\documentclass[a4paper,11pt,english]{article}

\usepackage{geometry}
\usepackage{lineno}

\usepackage{amsthm}
\usepackage{lmodern}
\usepackage{amsmath,amssymb,bm}
\usepackage{mathtools}
\usepackage{graphicx}
\usepackage{epstopdf}
\usepackage{latexsym}
\usepackage[normalem]{ulem} 
\usepackage{appendix}
\usepackage{dsfont}
\usepackage[usenames,dvipsnames]{color}

\usepackage{prettyref}
\usepackage[numbers,sort&compress]{natbib}

\usepackage[dvipsnames]{xcolor}
\usepackage{braket}
\usepackage{simpler-wick}
\usepackage{empheq}

\usepackage{environ}
\NewEnviron{equationS}{%
\begin{align}\begin{split}
  \BODY
\end{split}\end{align}
}
\definecolor{darkgreen}{rgb}{0,0.5,0}
\definecolor{darkblue}{rgb}{0,0,0.6}
\definecolor{purple}{rgb}{0.4,.2,0.7}
\usepackage[colorlinks=true,citecolor=magenta,linkcolor= blue,urlcolor=purple]{hyperref}

\numberwithin{equation}{section}
\numberwithin{figure}{section}
\numberwithin{table}{section}

\DeclareFontShape{OT1}{cmr}{mx}{n}{<->cmr10}{}

\begin{document}

\fontseries{mx}\selectfont

\newcommand{\vpd}[0]{\vphantom{\dagger}}
\newcommand{\vps}[0]{\vphantom{*}}
\newcommand{\vpp}[0]{\vphantom{\prime}}

\newcommand{\Rev }[1]{{\color{blue}{#1}\normalcolor}} % Revision
\newcommand{\Com}[1]{{\color{red}{#1}\normalcolor}} %Comment

\newcommand{\ketbra}[2]{|#1\rangle\langle #2|}%ketbra
\newcommand{\normord}[1]{\mathopen{:}\,#1\,\mathopen{:}}
\newcommand{\kett}[1]{|#1\rangle\hspace{-0.075cm}\rangle}
\newcommand{\bbra}[1]{\langle\hspace{-0.075cm}\langle #1|}
\newcommand{\bbrakett}[1]{\langle\hspace{-0.075cm}\langle #1 \rangle\hspace{-0.075cm}\rangle}
\newcommand{\brakett}[1]{\langle #1 \rangle\hspace{-0.075cm}\rangle}
\newcommand{\bbraket}[1]{\langle\hspace{-0.075cm}\langle #1 \rangle}
\newcommand{\bbrakettPsi}[1]{\langle\hspace{-0.075cm}\langle \hat{\Psi}|#1|\hat{\Psi} \rangle\hspace{-0.075cm}\rangle}
\def\normOrd#1{\mathop{:}\nolimits\!#1\!\mathop{:}\nolimits}

%\title{Large N vector models in the Hamiltonian framework}

%\author{D. Barberena}
%\it{JILA, Department of Physics, University of Colorado,  Boulder, CO 80309, USA}
%\it{Center for Theory of Quantum Matter, University of Colorado, Boulder, CO 80309, USA}
%\it{T.C.M. Group, Cavendish Laboratory, University of Cambridge, J.J. Thomson Avenue, Cambridge CB3 0HE, UK}
%\emailAdd{db985@cam.ac.uk}
\begin{center}
\LARGE \bf Large N vector models in the Hamiltonian framework
\end{center}

\vskip1cm

\begin{center}
Diego Barberena$^{1,2,3}$
\vskip5mm
{\it{\footnotesize $^1$ JILA, Department of Physics, University of Colorado,  Boulder, CO 80309, USA} \\
\it{\footnotesize $^2$ Center for Theory of Quantum Matter, University of Colorado, Boulder, CO 80309, USA}\\
\it{\footnotesize $^3$ T.C.M. Group, Cavendish Laboratory, University of Cambridge, J.J. Thomson Avenue, Cambridge CB3 0HE, UK}\\}
\end{center}

\vskip10mm
{\bf \center Abstract\\}
\vskip8mm
{We present a fluctuating $N$ formalism, based on second-quantization, to describe large $N$ vector models from field theory using Hamiltonian methods. We first present the method in the simpler setting of a quantum mechanical system with quartic interactions, and then apply these techniques to the $O(N)$ model in $2+1$ and $3+1$ dimensions. We recover various known results, such as the gap equation determining the ground state of the system, the presence of bound states at negative coupling and the leading order contribution to critical exponents, and provide an interpretation of the large $N$ path integral saddle point as a Bose-Einstein condensate of extended objects in the presence of a non-local interaction. In the large $N$ limit, this formalism leads naturally to a description of elementary $O(N)$ symmetric excitations in terms of bilocal fields, which are at the basis of $\text{AdS}_4/\text{CFT}_3$ studies of the $O(N)$ model and Vasiliev gravity. }

\newpage

\tableofcontents{}
\section{Introduction}
The large $N$ approximation is, by now, a standard part of the toolbox used to analyze quantum field theories. It affords various simplifications, sometimes rendering a theory solvable, and generally restricting the set of processes that can contribute to a given observable~\cite{klebanov2018tasi,HOOFT1974}. At the same time, the large $N$ setting is important in studies of holography, where $1/N$ controls the quantum effects of the gravity dual of the field theory under study~\cite{Maldacena1999,Witten1998,KLEBANOV2002213}. 

Calculations at large $N$ are typically done diagrammatically, where the terms in a perturbative expansion are grouped according to their $N-$dependence and then resummed order-by-order in $N$. In the case of vector models~\cite{MOSHE2003}, the approximation can also be implemented at the level of path integrals by introducing decoupling fields via a Hubbard-Stratonovich transformation and doing a saddle-point approximation. This provides access to the leading order term in $1/N$ to all orders in coupling constants, and higher order corrections can then be computed systematically.

%. We believe this approach has been relatively unexplored and we find that it provides a different, possibly useful, interpretation of the vector large $N$ approximation
The solvability of large $N$ vector models suggests that they should be amenable to analysis from complementary perspectives. In this paper we describe one such perspective based on the Hamiltonian framework, in which $N$ is allowed to be a fluctuating quantity, as originally presented in Ref.~\cite{Maslov199}. This Hamiltonian formalism is set-up in close analogy to the passage from first to second-quantization in many-body quantum theory~\cite{Altland_Simons_2010}, a context in which $N$ stands for the number of particles. Similarly, large $N$ vector models in $d+1$ dimensions can be interpreted as theories of $N$ objects that extend over $d$ spatial dimensions. From this point of view, the large $N$ saddle invoked in the path integral formalism corresponds to a Bose-Einstein condensate of these objects. Excitations about the condensate can then be studied using standard techniques. While the system possesses many kinds of excitations, a prominent role is played by those that naturally exhibit a bi-local structure, as originally found in Refs.~\cite{Maslov199,Shvedov2001}. This provides a direct connection to investigations of holography in vector models~\cite{KLEBANOV2002213,Das2003}, where derivations of the $\text{AdS}_4/\text{CFT}_3$ correspondence rely on collective bilocal fields~\cite{Das2003} on the field theory side~\cite{Aharony2021,Aharony2023b,deMelloKoch2023}. Furthermore, Hamiltonian formulations of field theories are useful because they are often the starting point for the application of a variety of techniques such as tensor-networks~\cite{Dalmonte02072016}, semi-definite programming~\cite{berenstein2021bootstrappingsimpleqmsystems,Lawrence2023}, and quantum simulation~\cite{halimeh2023coldatomquantumsimulatorsgauge,Bauer2023}.

The Hamiltonian point of view, in which quantum states play a center role, provides a very familiar picture of some of the low energy states of the large $N$ system. The ground state of the theory is, to leading order, a ``macroscopically populated" state of the $N=1$, ``single-particle", system. Furthermore, this ``single-particle" state is determined via a  nonlinear equation, in clear analogy to how the nonlinear Gross-Pitaevskii equation determines the condensate wavefunction of a many-body quantum system~\cite{2006Leggett}. %The solution to this nonlinear equation is non-unique, and all the solutions that are not the ground state describe highly energetic (with energies $\propto N$) states of the theory. 
Excited states are then described by a collection of independent harmonic oscillators, whose frequencies encode both bound and scattering state energies. Further corrections in $1/N$ provide interactions between these oscillators and hence, e.g., between bound states.

The application of the fluctuating $N$ formalism to multicomponent field theories has been considered before~\cite{Maslov199,Shvedov2001}, with an emphasis on the semiclassical nature of the large $N$ limit, and is also used in quantum cosmology under the name of third-quantization~\cite{RUBAKOV1988503,RoblesPerez2010} (not to be confused with the third-quantization technique used in open quantum systems~\cite{Prosen_2008}). Here we provide: (i) a formulation of the method in a basis-independent way that highlights the connections to Bose-Einstein condensation, (ii) a discussion of $O(N)$ invariance within the formalism, and (iii) some brief comments on the extensions to finite temperature and fermion models. As in the path integral formalism, this Hamiltonian framework also allows for a systematic expansion in powers of $1/N$.

This paper is organized as follows: 
\begin{description}
    \item[Section~\ref{sec:QuantumModel}] We focus on the simpler context of $0+1$d quantum mechanics to illustrate the technical aspects of the method and to introduce notation that will simplify formulas when we move on to field theory. We also discuss the $O(N)$ invariance constraint within the formalism and the conditions that it imposes on the spectrum of the system.
    \item[Section~\ref{sec:O(N)model}] We apply these techniques to the $O(N)$ model in $2+1$ and $3+1$ dimensions formulated on the lattice. At leading order we recover and solve the gap equation that determines the various path integral saddles that have been studied in the literature~\cite{MOSHE2003}. In light of the recent discussions about the role of $\mathcal{PT}-$symmetry~\cite{Ai2022,Romatschke2023,romatschke2023quantum} in the $O(N)$ model, we include also the results at negative coupling. At next to leading order we study the excitations of the system, which include a bound state at negative coupling, as well as the correction to the ground state energy, two-point correlators and critical exponents. 
    \item[Section~\ref{sec:FiniteTAndFermions}] We briefly consider finite temperature systems, indicating the differences that arise because of the imposition of permutational symmetry. We find that, if we restrict to $O(N)$ invariant states and take seriously the interpretation in terms of extended objects, there is a crossover from condensation to non-condensation at a finite temperature. However, this crossover is not reflected in the free energy of the system. We also provide a few comments on fermionic models such as Gross-Neveu~\cite{Gross1974}.
\end{description}

\section{Quantum Mechanical Model}\label{sec:QuantumModel}
Here we will consider the following interacting Hamiltonian:
\begin{equation}\label{eqn:QMHami}
    \hat{H}_N=\sum_{\alpha=1}^N\hat{p}^2_{\alpha}+\frac{\lambda}{N}\bigg(\sum_{\alpha=1}^N\hat{x}_\alpha^2\bigg)^2,
\end{equation}
which is an $N$ component generalization of the quartic oscillator ($N=1$) and whose properties have been heavily analyzed in the past~\cite{Ferrell1974,Bray_1974,Dolgov19880,Koudinov1982}. Going forward, we will refer to any of these $N$ components as a ``species". The operators $\hat{x}_\alpha$ and $\hat{p}_\beta$ form canonical pairs, i.e. $[\hat{x}_\alpha,\hat{p}_\beta]=i\delta_{\alpha\beta}$, and they act on quantum states that live in the tensor product Hilbert space $\mathcal{H}_N=\mathcal{H}^{\otimes N}$, where $\mathcal{H}$ is the Hilbert space of the $N=1$ system. We also define $\mathcal{H}_N$ for $N=0$ to be one-dimensional, with basis state $\ket{\text{vac}}$ acting as a vaccuum in ``species space". We are indexing the Hamiltonians $\hat{H}_N$ with $N$ because we will shortly consider all $N$ simultaneously.

In the way the problem is framed, $N$ is fixed and the operators $\hat{x}_\alpha,\hat{p}_\alpha$ act on the Hilbert space at this fixed value of $N$. However, given that the Hamiltonian in Eq.~(\ref{eqn:QMHami}) is permutationally symmetric under exchange of species, the symmetric subspace of each $\mathcal{H}_N$ is left invariant by $\hat{H}_N$, and the eigenvalue problem in this sector can be reformulated in terms of operators that switch between different $N$. Namely, one replaces~\cite{Altland_Simons_2010}
\begin{equation}\label{eqn:QMChangeOfRep}
    \sum_{\alpha=1}^N\hat{x}_{\alpha}^2=\int\, x^2\hat{\Psi}_x^\dagger\hat{\Psi}_x\,dx,\hspace{1cm}
    \sum_{\alpha=1}^N\hat{p}_{\alpha}^2=-\int \hat{\Psi}_x^\dagger\partial_x^2\hat{\Psi}_x\,dx,
\end{equation}
\begin{figure}
    \centering
    \includegraphics[width=0.7\linewidth]{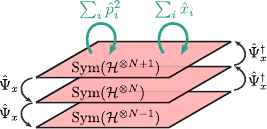}
    \caption{The operators $\hat{\Psi}_x$ and $\hat{\Psi}_x^\dagger$ map a state with $N$ species to a state with $N-1$ and $N+1$ species, respectively. Operators that are symmetric under permutations and that do not change the number of species, e.g. $\sum_i \hat{p}_i^2$, can be represented in terms of $\hat{\Psi}_x,\hat{\Psi}_x^\dagger$ when acting on permutationally symmetric states.}
    \label{fig:SecondQuantization}
\end{figure}
where $\left[\hat{\Psi}_{x\vphantom{'}}^{\vphantom{\dagger}},\hat{\Psi}_{x'}^\dagger\right]=\delta (x-x')$ are bosonic field operators that change the number of species (see Fig.~\ref{fig:SecondQuantization}): $\hat{\Psi}_x$ destroys one species ($\mathcal{H}_N\to\mathcal{H}_{N-1}$), while $\hat{\Psi}_x^\dagger$ creates one ($\mathcal{H}_N\to\mathcal{H}_{N+1}$). In terms of these field operators, the set of Hamiltonians $\{\hat{H}_N\}$ can be rewritten as
\begin{equation}\label{eqn:QMHami2}
    \hat{H}=-\frac{1}{2}\int\,dx \hat{\Psi}_x^\dagger\partial_x^2\hat{\Psi}_x+\frac{\lambda}{\hat{N}}\bigg(\int\, x^2\hat{\Psi}_x^\dagger\hat{\Psi}_x\,dx\bigg)^2,
\end{equation}
and $\hat{N}$ is now an operator too. The leading order approximation (LO) in $1/N$ is obtained by replacing $\hat{\Psi}_x$ by a $c$-number, as in the theory of Bose-Einstein condensation~\cite{2006Leggett}. Before heading down that route, let us introduce some notation. First, Eqs.~(\ref{eqn:QMChangeOfRep}) are specific instances of the more general formula~\cite{Altland_Simons_2010}
\begin{equation}
    \sum_{\alpha=1}^N\hat{O}_\alpha=\sum_{n,m}\braket{m|\hat{O}|n}\hat{\Psi}^\dagger_{m}\hat{\Psi}_n,
\end{equation}
where $\hat{O}$ is any operator that acts on $\mathcal{H}$, the set $\{\ket{n}\}$ is a basis of $\mathcal{H}$, $[\hat{\Psi}_{n}^{\vpd},\hat{\Psi}_{m}^\dagger]=\delta_{nm}$, and $\hat{\Psi}_{n}^\dagger$ can be interpreted as an operator that ``creates a species" in state $\ket{n}$. We can see that Eq.~(\ref{eqn:QMChangeOfRep}) is recovered when $\ket{n}$ are taken to be eigenstates of $\hat{x}$ and $\hat{O}=\hat{x}^2\text{ or }\hat{p}^2$. To avoid making a reference to any specific basis of $\mathcal{H}$, we introduce the abstract basis-independent object
\begin{equation}
    \kett{\hat{\Psi}}=\sum_{n}\ket{n}\hat{\Psi}_n,
\end{equation}
which involves both operators $\hat{\Psi}_n$ and ``states" $\ket{n}\in\mathcal{H}\neq \bigoplus_N\text{Sym}(\mathcal{H}_N)$. Importantly, $\hat{\Psi}_n$ does not act on $\ket{n}\in \mathcal{H}$, but instead on the true multi-species Hilbert space of the system $\bigoplus_N\text{Sym}(\mathcal{H}_N)$. In this sense, $\ket{n}$ are just a mathematical device that allows us to formulate the problem in a basis-independent way. Alternatively, $\ket{n}$ can be thought of as ``single particle" or (more precisely) ``single species" states, and basis states of $\bigoplus_N\text{Sym}(\mathcal{H}_N)$ can be described by giving  the occupation numbers of the different $\ket{n}$. To avoid confusion, we will refer to $\ket{n}\in\mathcal{H}$ as $sp-$states, while the true quantum states $\in \bigoplus_N\text{Sym}(\mathcal{H}_N)$ will be referred to just as states. In terms of the object $\kett{\hat{\Psi}}$, Eq.~(\ref{eqn:QMHami2}) can be written succinctly as
\begin{equation}\label{eqn:QMHami3}
    \hat{H}=\frac{1}{2}\bbrakett{\hat{\Psi}|\,\hat{p}^2|\hat{\Psi}}+\frac{\lambda}{\hat{N}} \bbrakett{\hat{\Psi}|\,\hat{x}^2|\hat{\Psi}}^2.
\end{equation}
The ``species" number $\hat{N}$ also admits the following simple representation:
\begin{equation}\label{eqn:QMN}
    \hat{N}=\bbrakett{\hat{\Psi}|\hat{\Psi}}.
\end{equation}
The standard interpretation adscribed to this formalism is that the variables $\hat{x}_{\alpha}$ are the positions of $N$ zero-dimensional objects along a one dimensional line (with coordinate given by $x$). This is thus a description of $N$ bosonic particles, with a standard kinetic term and a non-local, non-translationally invariant (in the direction of the $\hat{x}$) interaction.\\
The sectors of $\mathcal{H}_N$ that are antisymmetric under the exchange of species are also accessible by these methods. The only modification is that the set of $\hat{\Psi}_n$ will now be fermionic and will satisfy $\{\hat{\Psi}_n,\hat{\Psi}_{n'}\}=\delta_{nn'}$. Furthermore, to obtain a non-trivial large $N$ limit the scaling of $\lambda$ with $N$ in Eq.~(\ref{eqn:QMHami}) needs to be modifed ($\lambda/N\to \lambda/N^2$). We leave the study of these states for another paper.
\subsection{Leading order (LO)}\label{subsec:QMLeadingOrder}

The leading order approximation is obtained by assuming that only one of the $sp-$states $\ket{n}$ becomes ``macroscopically populated". We denote this $sp-$state as $\ket{F}$ and the associated creation operator as $\hat{\Psi}^\dagger_F\equiv \hat{F}^\dagger$. Then $\kett{\hat{\Psi}}\approx \ket{F}\hat{F}$ (we are omitting the other non macroscopically populated $sp-$states), while
\begin{equationS}
    \hat{N}&\approx \hat{F}^\dagger\hat{F}\\
    \hat{H}_{\text{LO}}&\approx \hat{N}\bigg[\frac{1}{2}\braket{F|\hat{p}^2|F}+\lambda \braket{F|\hat{x}^2|F}^2\bigg].
\end{equationS}
To leading order, the Hamiltonian is the operator $\hat{N}$ times a $c$-number that depends non-linearly on the $sp-$state $\ket{F}$. We now minimize this $c$-number over all $sp-$states $\ket{F}$ subject to the condition $\braket{F|F}=1$. This leads to the eigenvalue equation
\begin{equation}\label{eqn:QMSelfConsistency}
    \underbrace{\bigg(\frac{\hat{p}^2}{2}+2\lambda\,\sigma\,\hat{x}^2\bigg)}_{\hat{h}_{\sigma}}\ket{F}=\mu\ket{F},
\end{equation}
where $\sigma\equiv\braket{F|\,\hat{x}^2|F}$, $\mu$ is a Lagrange multiplier and $\hat{h}_{\sigma}$ is a ``single species" harmonic oscillator Hamiltonian whose frequency $\omega_\sigma^2=4\lambda\sigma$ needs to be determined self-consistently. This is a nonlinear equation because of the self-consistency condition and is equivalent to the nonlinear Gross-Pitaevskii equation that determines the condensate wavefunction of a BEC (namely, which single particle state is macroscopically populated). We solve Eq.~(\ref{eqn:QMSelfConsistency}) by introducing the $\sigma$ dependent annihilation operator 
\begin{equation}\label{eqn:QMannihilation}
    \hat{a}_{\sigma}=\frac{\omega_{\sigma}\hat{x}+i\hat{p}}{\sqrt{2\omega_\sigma}}.
\end{equation}
We now have that $\hat{h}_{\sigma}=\omega_{\sigma}(\hat{a}^\dagger_{\sigma}\hat{a}_{\sigma}^{\vpd}+1/2)$. We take $\ket{F}$ to be the harmonic oscillator $sp-$eigenstate with 0 excitations (i.e. $\hat{a}_{\sigma}\ket{F}=0$). Then we find that
\begin{equationS}\label{eqn:QMLOSolutions}
    \sigma_*&=\frac{1}{2(2\lambda)^{1/3}}\longrightarrow \omega_{\sigma_*}=(2\lambda)^{1/3}\\[5pt]
    \hat{H}_{\text{LO}}&=\hat{N}\bigg[\frac{3}{8}(2\lambda)^{1/3}\bigg]\approx 0.375(2\lambda)^{1/3}\hat{N},
\end{equationS}
This result is in agreement with Refs.~\cite{Hioe1978,romatschke2023quantum}. At this level of approximation, the ground state of the system is
\begin{equation}
    \ket{\text{GND}}_{\text{LO}}=\frac{(\hat{F}^\dagger)^N}{\sqrt{N!}}\ket{\text{vac}}=\ket{F}^{\otimes N},
\end{equation}
In the previous equation, the term in the middle is an expression that employs operators that change $N$, while the term on the right is meaningful also at fixed $N$. Eq.~(\ref{eqn:QMSelfConsistency}) has other solutions $\ket{F'}$ that make $\hat{H}_{\text{LO}}$ stationary with respect to infinitesimal variations in the $sp-$state, but lead to higher energy states, with energy gaps $\propto N$. Because of the self-consistency condition for $\sigma$, they are $sp-$eigenstates of harmonic oscillators with different frequencies. The immediate consequence of this is that these $sp-$solutions are not orthogonal among each other. Nevertheless the overlaps of their associated quantum states (which live in the multispecies Hilbert space) are exponentially small in $N$ (e.g. $\braket{F|F'}^N$). We defer a discussion of these other solutions to Appendix~\ref{app:QMStationaryPoints} (see also Ref.~\cite{Maslov199}). There exist even more leading order solutions, corresponding to fragmented BEC's~\cite{2006Leggett}, where more than one $sp-$state is macroscopically occupied (e.g. $\ket{\text{GND}}_{\text{LO}}\propto (\hat{F}_1^{\dagger})^{N_1}(\hat{F}_2^{\dagger})^{N_2}\ket{\text{vac}}$, with $N_1+N_2=N$). In a similar fashion, the nature of the associated $sp-$states needs to be determined self-consistently.
\subsubsection{Expectation values}
Expectation values of permutationally symmetric operators become simple at leading order. Replacing $\kett{\hat{\Psi}}\approx \hat{F}\ket{F}$ and $\hat{F}^\dagger\hat{F}\approx \hat{N}$ we get
\begin{equation}\label{eqn:QMCorrelatorLO}
    \hat{D}\equiv\sum_{\alpha}\hat{x}_\alpha^2=\bbrakettPsi{\hat{x}^2}\approx\hat{N}\braket{F|\,\hat{x}^2|F}=\frac{\hat{N}}{2(2\lambda)^{1/3}},
\end{equation}
which is just a c-number times $\hat{N}$. Because of this, products of operators factorize
\begin{equation}
    \hat{D}^2=\left(\sum_{\alpha}\hat{x}_\alpha^2\right)\left(\sum_{\alpha}\hat{x}_\alpha^2\right)=\hat{N}^2\braket{F|\,\hat{x}^2|F}^2=\frac{\hat{N}^2}{4(2\lambda)^{2/3}},
\end{equation}
as is well known, and we expect quantum fluctuations to be of higher order in $1/\hat{N}$.
\subsection{Next to leading order (NLO)}

The leading order contribution to the energy was obtained by keeping only one $sp-$state in the expansion for $\kett{\hat{\Psi}}$. We go beyond LO by including all the $sp-$states, but keeping $\ket{F}$ separate:
\begin{equation}
    \kett{\hat{\Psi}}=\ket{F}\hat{F}+\underbrace{\sum_{n\neq F}\ket{n}\hat{\Psi}_n}_{\kett{\delta\hat{\Psi}}}.
\end{equation}
Importantly, since $\{\ket{n},\ket{F}\}$ form an orthonormal basis, we have that $\brakett{F|\delta\hat{\Psi}}=0$. The number of species can be expressed in terms of this decomposition as
\begin{equation}\label{eqn:QMNLOHami}
    \hat{N}=\hat{F}^\dagger\hat{F}+\bbrakett{\delta\hat{\Psi}|\delta\hat{\Psi}},
\end{equation}
and the same thing can be done for the Hamiltonian $\hat{H}$ in Eq.~(\ref{eqn:QMHami3}). After grouping terms according to their power of $N$ (see Appendix~\ref{app:QMNLOHami}) we find that the Hamiltonian to order $O(N^0)$ is
\begin{equationS}\label{eqn:QMNLOHamiltonian}
    \hat{H}_{\text{NLO}}&=0.375(2\lambda)^{1/3}\hat{N}+\bbra{\delta\hat{\Psi}}\big(\hat{h}_*-\mu\big)\kett{\delta\hat{\Psi}}+\lambda\left[e^{-i\hat{\varphi}}\brakett{F|\hat{x}^2|\delta\hat{\Psi}}+\text{h.c.}\right]^2.
\end{equationS}
The first term in Eq.~(\ref{eqn:QMNLOHamiltonian}) is the LO contribution. The second term involves $\hat{h}_{*}\equiv\hat{h}_{{\sigma}_*}$, which is the $sp-$hamiltonian evaluated at $\sigma_*$ [Eq.~(\ref{eqn:QMLOSolutions})] and $\mu$ is defined in Eq.~(\ref{eqn:QMSelfConsistency}). The third term uses the phase operator $e^{i\hat{\varphi}}=(\hat{F}^\dagger\hat{F}+1)^{-1/2}\,\hat{F}$, which reduces $\hat{F}^\dagger\hat{F}$ by one with unit amplitude~\cite{Susskind1964,Kromminga1962}. The object $\brakett{F|\hat{x}^2|\delta\hat{\Psi}}$ is also an annihilation operator, obtained by extracting the projection of $\kett{\delta\hat{\Psi}}$ onto the $sp-$state $\hat{x}^2\ket{F}$. This is a basis-independent expression, but a convenient choice of $sp-$basis will simplify both this operator and the NLO Hamiltonian. Note that all the new terms [as compared to Eq.~(\ref{eqn:QMLOSolutions})] contain two $\delta\hat{\Psi}$ operators. Terms linear in $\delta\hat{\Psi}$, which would be of size $\sim N^{1/2}$, disappear by virtue of the stationarity condition of $\ket{F}$ (see Appendix~\ref{app:QMNLOHami}).

The structure of $\hat{H}$ indicates that it is convenient to expand $\kett{\delta\hat{\Psi}}$ in terms of $\{\ket{n}\}$ that are $sp-$eigenstates of $(\hat{h}_{*}-\mu)\propto\hat{a}^{\dagger}_{\sigma_*}\hat{a}_{\sigma_*}^{\vpd}$. To avoid confusion with the generic label $n$, we denote the $\nu-$th level of $\hat{h}_{*}$ as $\ket{\nu}\propto (\hat{a}_{\sigma_*}^{\dagger})^{\nu}\ket{F}$, so that $\kett{\delta\hat{\Psi}}=\sum_{\nu=1}\ket{\nu}\hat{\Psi}_\nu$. We can thus rewrite the Hamiltonian as
\begin{equationS}\label{eqn:QMNLODiagonal}
    \hat{H}_{\text{NLO}}&=(2\lambda)^{1/3}\left[0.375\hat{N}+\sum_{\nu=1}^{\infty}\nu\,\Psi_{\nu}^{\dagger}\Psi_{\nu}+\frac{1}{4}\Big(\hat{\Psi}_2^\dagger e^{i\hat{\varphi}}+\hat{\Psi}_2 e^{-i\hat{\varphi}}\Big)^2\right].
\end{equationS}
In the multispecies Hilbert space, elementary excitations are labeled ($\nu\geq 1$) by the $sp-$eigenstates of $\hat{h}_{*}$ orthogonal to $\ket{F}$ (hence $\nu=0$ does not appear in the sum). In the ground state of the system, all $\Psi^\dagger_{\nu}\Psi_{\nu}$ are 0 except for $\nu=2$, which is modified due to the interactions. Importantly, this Hamiltonian still conserves the number of species $N$ by virtue of the operator $e^{i\hat{\varphi}}$.

The Hamiltonian for the $\nu=2$ sector is gaussian, and can be diagonalized by means of a Bogoliubov transformation. It contributes a constant piece to the ground state energy
\begin{equation}
    E_{\text{NLO}}^{\text{GND}}=(2\lambda)^{1/3}\left[0.375 N+\bigg(\sqrt{\frac{3}{2}}-1\bigg)\right],
\end{equation}
which agrees with Ref.~\cite{romatschke2023quantum}. At NLO, the ground state of the system is now
\begin{equation}\label{eqn:QMNLOGstate}
    \ket{\text{GND}}_{\text{NLO}}=\frac{1}{(1-z^2)^{-1/4}}\exp\bigg[\frac{z}{2}(\hat{\Psi}_2^\dagger)^2e^{2i\hat{\varphi}}\bigg]\frac{(F^\dagger)^{N}}{\sqrt{N!}}\ket{\text{vac}}
\end{equation}
where $z=-1/(5+\sqrt{24})\approx-0.101$ is a pure number. The occupation of the $sp-$state $\ket{F}$ is no longer equal to $N$, and there is a finite number of excitations $\hat{\Psi}_2^\dagger$, which appear at expense of depleting condensate. The condensate depletion in the ground state is
\begin{equation}
    \left\langle\hat{N}-\hat{F}^\dagger\hat{F}\right\rangle=\braket{\hat{\Psi}_2^\dagger\hat{\Psi}_2}=\frac{1}{4\sqrt{6}}\left(\frac{1}{5+\sqrt{24}}\right)\approx 0.01,
\end{equation}
and is small compared to the condensate occupation $\sim N$.
% The state can also be written without resorting to operators that change $N$, but the expression is not particularly illuminating, so we relegate it to Appendix~\ref{app:QMGroundState}. 
Further corrections are obtained by keeping higher orders in the $1/N$ expansion in the Hamiltonian and applying standard perturbation theory. As shown in Appendix~\ref{app:QMNLOHami}, the Hamiltonian includes corrections to all orders in $1/\sqrt{N}$.

\subsubsection{Correlators}
We can calculate corrections to expectation values by applying the decomposition $\kett{\hat{\Psi}}=\hat{F}\ket{F}+\kett{\delta\hat{\Psi}}$ to operators like $\hat{D}=\bbrakettPsi{\,\hat{x}^2}$ in Eq.~(\ref{eqn:QMCorrelatorLO}). For example, 
\begin{equation}
    \hat{D}=\hat{F}^\dagger\hat{F}\braket{F|\,\hat{x}^2|F}+\left[\hat{F}^\dagger\brakett{F|\,\hat{x}^2|\delta\hat{\Psi}}+\bbraket{\delta\hat{\Psi}|\,\hat{x}^2|F}\hat{F}\right]+\bbrakett{\delta\hat{\Psi}|\,\hat{x}^2|\delta\hat{\Psi}}.
\end{equation}
Using the replacements $\hat{F}^\dagger\hat{F}=\hat{N}-\bbrakett{\delta\hat{\Psi}|\delta\hat{\Psi}}$ and $\hat{F}\approx e^{i\hat{\varphi}}\hat{N}^{1/2}$, we can rewrite this expression as
\begin{equation}\label{eqn:QMLeadingOrderCorre}
    \hat{D}=\frac{\hat{N}}{2(2\lambda)^{1/3}}+\frac{\hat{N}^{1/2}}{(2\lambda)^{1/3}}\left(\frac{\hat{\Psi}_2e^{-i\hat{\varphi}}+\hat{\Psi}_2^{\dagger}e^{i\hat{\varphi}}}{\sqrt{2}}\right)+O(\hat{N}^0),
\end{equation}
which includes contributions only up to $\sim \hat{N}^{1/2}$. From this last expression we can do two things:
\begin{enumerate}
    \item Define the fluctuation operator
    \begin{equation}
        \delta\hat{D}=\bbrakettPsi{\,\hat{x}^2}-\frac{\hat{N}}{2(2\lambda)^{1/3}},
    \end{equation}
    and calculate its second order correlator
    \begin{equation}
        \braket{\delta\hat{D}^2}=\frac{N}{(2\lambda)^{2/3}}\left\langle\left(\frac{\hat{\Psi}_2 e^{-i\hat{\varphi}}+\hat{\Psi}_2^{\dagger}e^{i\hat{\varphi}}}{\sqrt{2}}\right)^2\right\rangle=\frac{N}{(2\lambda)^{2/3}}\left(\frac{1}{\sqrt{6}}\right),
    \end{equation}
    which can be interpreted as the size of fluctuations of $\bbrakettPsi{\,\hat{x}^2}$ about its mean value. Schematically, $\delta\hat{D}\sim N^{1/2}$. This required us to keep only the $\hat{N}^{1/2}$ term in Eq.~(\ref{eqn:QMLeadingOrderCorre}).
    \item Calculate direct corrections to $\braket{\hat{D}}$. The $\hat{N}^{1/2}$ contribution vanishes by symmetry so we should consider the $\hat{N}^0$ correction. To calculate this, we need $\hat{D}$ to $O(\hat{N}^0)$. Unfortunately, the $O(N^{1/2})$ term in Eq.~(\ref{eqn:QMLeadingOrderCorre}) will pair up with NNLO corrections to the ground state (which are of size $\hat{N}^{-1/2}$) to contribute to order $\hat{N}^0$ in $\braket{\hat{D}}$ and thus requires further calculations to determine the NNLO wavefunction. Fortunately, this complication is absent for certain operators. For instance, for
    \begin{equationS}
        \hat{\Upsilon}\equiv\bbrakettPsi{\hat{h}_{\sigma_*}}&=\hat{N}\braket{F|\hat{h}_{\sigma_*}|F}+\hat{N}^0\left[\bbrakett{\delta\hat{\Psi}|\hat{h}_{\sigma_*}|\delta\hat{\Psi}}-\braket{F|\hat{h}_{\sigma_*}|F}\right]
    \end{equationS}
    the $\hat{N}^{1/2}$ term vanishes by virtue of the fact that $\ket{F}$ is a $sp-$eigenstate of $\hat{h}_{\sigma_*}$ [defined in Eq.~(\ref{eqn:QMSelfConsistency})]. In consequence, to calculate this operator to $O(\hat{N}^0)$ only the NLO wavefunction suffices:
    \begin{equation}
    \braket{\hat{\Upsilon}}=(2\lambda)^{1/3}\left(\frac{N}{2}+2\braket{\hat{\Psi}_2^\dagger\hat{\Psi}_2}\right)=(2\lambda)^{1/3}\left(\frac{N}{2}+\frac{5}{2\sqrt{6}}-1\right)
    \end{equation}
    Furthermore, in contrast to $\hat{D}$, whose fluctuations are of size $\delta\hat{D}\sim N^{1/2}$, the fluctuations of $\hat{\Upsilon}$ are suppressed and  of size $\delta\hat{\Upsilon}\sim N^0$. The ability to perform this choice will simplify the calculation of critical exponents in the field theory context.
\end{enumerate}

\subsection{O(N) invariant states}\label{sec:QMONInvariance}
We finalize the section on the quantum mechanical model by discussing $O(N)$ invariance. By construction, the different excitations of $\hat{H}$ are permutationally invariant. However it is possible to further restrict the spectrum of the system by requiring $O(N)$ invariance. This is a condition that needs to be imposed on quantum states\footnote{Naturally, this can be done in the fixed $N$ formulation by requiring the wavefunction to depend only on $\sum_\alpha x_\alpha^2$, but here we seek to recast this condition when we allow $N$ to fluctuate.}. In the fixed $N$ formulation, rotations between species $\alpha$ and $\beta$ are generated by the operator
\begin{equation}
    \hat{x}_\alpha\hat{p}_\beta-\hat{x}_\beta\hat{p}_\alpha,
\end{equation}
which has eigenvalue $0$ in $O(N)$ invariant states. The permutationally symmetric way of representing this condition is
\begin{equation}
    \sum_{\alpha\neq\beta}\left(\hat{x}_\alpha\hat{p}_\beta-\hat{x}_\beta\hat{p}_\alpha\right)\left(\hat{x}_\alpha\hat{p}_\beta-\hat{x}_\beta\hat{p}_\alpha\right)\ket{S}=0,
\end{equation}
where $\ket{S}$ is any $O(N)$ invariant state. The double sum over $\alpha,\beta$ means that this constraint is better interpreted as a two-body operator. From the rules of second quantization~\cite{Altland_Simons_2010}, in the fluctuating $N$ description this is represented as
\begin{equation}
   \left[ \bbra{\hat{\Psi}}_a\otimes\bbra{\hat{\Psi}}_b\big(\hat{x}_a\hat{p}_b-\hat{x}_b\hat{p}_a\big)^2\kett{\hat{\Psi}}_a\otimes\kett{\hat{\Psi}}_b\right]\ket{S}=0
\end{equation}
where we have introduced 
\begin{equation}
    \kett{\hat{\Psi}}_a\otimes\kett{\hat{\Psi}}_b=\sum_{n,m}{\hat{\Psi}}_n\hat{\Psi}_m\ket{n}_a\otimes\ket{m}_b,
\end{equation}
as a two-body analog of $\kett{\hat{\Psi}}$, and the $a,b$ subscript indicates which $sp-$operator acts on which $sp-$state. Since the constraint operator is the product of
\begin{equation}
   \big(\hat{x}_a\hat{p}_b-\hat{x}_b\hat{p}_a\big)\kett{\hat{\Psi}}_a\otimes\kett{\hat{\Psi}}_b
\end{equation} 
times its complex conjugate, demanding that it be $0$ is equivalent to demanding that
\begin{equation}\label{eqn:QMONinvariance}
      \left[\big(\hat{x}_a\hat{p}_b-\hat{x}_b\hat{p}_a\big)\kett{\hat{\Psi}}_a\otimes\kett{\hat{\Psi}}_b\right]\ket{S}=0,
\end{equation}
which is more manageable. Projecting Eq.~(\ref{eqn:QMONinvariance}) onto $sp-$states $\bra{n}_a\otimes\bra{m}_b$ provides one constraint, so that there is in fact an infinite number of conditions (for each pair $n,m$) that $O(N)$ invariant states need to satisfy. In terms of the operators $\hat{\Psi}_{\nu}$ and to leading order in $1/N$, these conditions take the form (Appendix~\ref{app:QMConstraint}):
\begin{equation}
    \hat{\Psi}_{\nu}\ket{S}=0,\text{ for }\nu\neq 2,
\end{equation}
thus restricting drastically the spectrum of excitations. In particular, the ground state in Eq.~(\ref{eqn:QMNLOGstate}) is $O(N)$ invariant and so is any state obtained by acting with any number of $\hat{\Psi}_2^\dagger$. All the other excitations are excluded from the $O(N)$ invariant spectrum to leading order in $1/N$. 
\subsubsection{Other consequences of the O(N) constraint}\label{sec:HigherOrderQM}
The $O(N)$ condition can be used to constrain the form of higher order terms in the $1/N$ expansion of $O(N)$ invariant states. For example, by projecting Eq.~(\ref{eqn:QMONinvariance}) onto adequate $sp$-states, we obtain the following constraints
\begin{align}\begin{split}
    \hat{\Psi}_4\hat{F}&=\sqrt{\frac{3}{2}}\hat{\Psi}_2^2\\
     \hat{\Psi}_6\hat{F}&=\sqrt{\frac{5}{3}}\hat{\Psi}_2\hat{\Psi}_4\rightarrow \hat{\Psi}_6\hat{F}^2=\sqrt{\frac{5}{2}}\hat{\Psi}_2^3,
\end{split}\end{align}
and in general
\begin{equation}
    \hat{\Psi}_{2k}\hat{F}^{k-1}\propto \hat{\Psi}_2^{k}.
\end{equation}
This indicates that $\braket{\hat{\Psi}^\dagger_{2k}\hat{\Psi}_{2k}^{\vpd}}\sim 1/N^{k-1}$. We now expand the ground state in powers of $N^{-1/2}$
\begin{equation}
    \ket{\text{GND}}=\ket{G_0}+\frac{1}{\sqrt{N}}\ket{G_1}+\frac{1}{N}\ket{G_2}+O(N^{-3/2}),
\end{equation}
where $\ket{G_0}$ is the NLO ground state given in Eq.~(\ref{eqn:QMNLOGstate}), and impose the above constraints by force, as well as the normalization choice $\braket{G|G_0}=1$. Taking into account that $\hat{F}=e^{i\hat{\varphi}}\hat{N}^{1/2}+O(N^{-1/2})$, this leads to
\begin{equation}
    \hat{\Psi}_4\ket{G_1}=\sqrt{\frac{3}{2}}e^{-i\hat{\varphi}}\hat{\Psi}_2^2\ket{G_0},
\end{equation}
with $\hat{\Psi}_{\nu}\ket{G_1}=0$ for $\nu\neq 2,4$. From these relations, we immediately find that
\begin{equation}
    \braket{\hat{\Psi}_4^\dagger\hat{\Psi}_4^{\vpd}}=\frac{3}{2N}\braket{G_0|(\hat{\Psi}_2^\dagger)^2(\hat{\Psi}_2^{\vpd})^2|G_0}=\frac{3}{2N}(3v^4+v^2),
\end{equation}
where $v^2=\braket{G_0|\hat{\Psi}_2^\dagger\hat{\Psi}_2|G_0}\approx 0.01$. More generally,
\begin{equation}
    \ket{G_1}=\sqrt{\frac{3}{2}}\hat{\Psi}_4^\dagger\hat{\Psi}_2^2e^{-i\hat{\varphi}}\ket{G_0}+\ket{R_0},
\end{equation}
where $\ket{R_0}$ is fixed by the Hamiltonian, and satisfies $\braket{G_0|R_0}$ and $\hat{\Psi}_{\nu}\ket{R_0}=0$ for $\nu\neq 2$ (it only contains $\hat{\Psi}_2$ excitations).
Similarly, to lowest non-vanishing order,
\begin{equation}
    \braket{\hat{\Psi}_{2k}^\dagger\hat{\Psi}_{2k}^{\vpd}}=\frac{1}{N^{k-1}}\frac{(2k)!}{2^k\,k!^2}\braket{(\hat{\Psi}_2^{\dagger})^{k}(\hat{\Psi}_2^{\vpd})^k}+O(N^{k-2})
\end{equation}
% In fact, the constraints can be treated exactly, and we have in general that
% \begin{equation}\label{eqn:OnInvarianceExactCorrelator}
%     \braket{\hat{\Psi}^\dagger_{2k}\hat{\Psi}_{2k}^{\vpd}}=\frac{(2k)!}{k!^2 2^k}\left\langle e^{i\hat{\varphi}(k-1)}(\hat{\Psi}_2^{\dagger})^k\left[\frac{\hat{N}_0!}{(\hat{N}_0+k)!}\right]\hat{\Psi}_2^ke^{i\hat{\varphi}(k-1)}\right\rangle,
% \end{equation}
%where $\hat{N}_0=\hat{F}^\dagger\hat{F}$.
provides average values of $\hat{\Psi}_{2k}$ in terms of those of $\hat{\Psi}_2,\hat{F}$. 

\section{O(N) model}\label{sec:O(N)model}
We illustrate these methods in the context of field theory by applying them to the well-studied $O(N)$ model in $d+1$ dimensions. To avoid dealing with the technical complications introduced by renormalization, we solve the theory on the lattice and take the continuum limit only at the end. As before, the LO calculations will amount to solving a self-consistent quadratic problem for the $N=1$ $sp-$states, with an emerging correlation length that must be tuned to be much larger than the lattice spacing (set to $1$ from here on) for a continuum description to be aproppriate. The NLO corrections will describe gaussian excitations that will very naturally organize themselves in terms of a bilocal description, as is common in studies of holography for large $N$ vector models. 

We consider a straightforward discretization of the $O(N)$ model, with fields at a spatial position replaced by a quantum bosonic variable $\hat{\phi}_r$, second derivatives in time replaced by terms quadratic in conjugate momentum operators $\hat{p}_r$, and spatial derivatives replaced by lattice hopping terms. Taking this into consideration, we end up with the following lattice Hamiltonian: 
\begin{equation}\label{eqn:SuSchriefferHegger}
    \hat{H}^{O(N)}=\sum_{r}\sum_{\alpha=1}^N\frac{(\hat{p}_r^\alpha)^2}{2}+\frac{1}{2}\sum_{r,\delta r} \sum_{\alpha=1}^NJ_{\delta r}^{\vphantom{\alpha}}\hat{\phi}_r^\alpha\hat{\phi}_{r+\delta r}^\alpha+\frac{\lambda}{N}\sum_r\left[\frac{1}{N}\sum_{\alpha=1}^N(\hat{\phi}_r^\alpha)^2\right]^2,
\end{equation}
Let us go over this expression one part at a time. First, $\alpha$ still denotes the different species and ranges from $1$ to $N$. Second, $r$ indexes position in a $d-$dimensional lattice with $V=L^d$ lattice sites, and $\delta r$ are adjacent directions to a given position $r$. Third, the degrees of freedom are bosonic and satisfy canonical commutation relations
\begin{equation}
    [\hat{\phi}_{r}^\alpha,\hat{p}_{r'}^\beta]=i\delta_{\alpha\beta}\delta_{r,r'},
\end{equation}
while all other commutators vanish. The second term in Eq.~(\ref{eqn:SuSchriefferHegger}) leads to the gradient terms in the continuum $O(N)$ model and we must keep track of the fact that we are choosing the lattice spacing to be $1$. We now rewrite $\hat{H}^{O(N)}$ in terms of $\kett{\hat{\Psi}}$:
\begin{equationS}\label{eqn:ONHamiltonianSecondQuantized}
    \hat{H}^{O(N)}&=\frac{1}{2}\sum_{r}\bbra{\hat{\Psi}}\hat{p}_r^2\kett{\hat{\Psi}}+\frac{1}{2}\sum_{r,\delta r}J_{\delta r}\bbrakett{\hat{\Psi}|\hat{\phi}_r\hat{\phi}_{r+\delta r}|\hat{\Psi}}+\frac{\lambda}{\hat{N}}\sum_{r}\Big[\bbra{\hat{\Psi}}\hat{\phi}_r^2\kett{\hat{\Psi}}\Big]^2,
\end{equationS}
% \begin{widetext}
%     \begin{equation}
%         \hat{H}^{\text{GN}}=J\sum_{r=1}^L\bbra{\hat{\Psi}}\Big(\hat{b}^\dagger_{r}\hat{a}_{r+1}+\hat{a}^\dagger_{r+1}\hat{b}_{r}-\hat{a}_{r}^\dagger\hat{b}_{r}-\hat{b}_{r}^\dagger\hat{a}_{r}\Big)\kett{\hat{\Psi}}-\frac{g^2}{\hat{N}}\sum_{r=1}^L\Big[\bbra{\hat{\Psi}}\Big(\hat{a}^\dagger_r\hat{a}_r-\hat{b}^\dagger_r\hat{b}_r\Big)\kett{\hat{\Psi}}\Big]^2
%     \end{equation}
% \end{widetext}
where $\hat{p}_r$ and $\hat{\phi}_r$ are $N=1$ operators that act on $sp-$states. As in the quantum mechanical case, this formalism lends itself to a ``multi-particle" interpretation: Eq.~(\ref{eqn:ONHamiltonianSecondQuantized}) describes a collection of $N$ bosonic objects that extend over $d-$dimensions, with coordinates $r$, that can move along a one dimensional line, with coordinate $\hat{\phi}$. The $\hat{\phi}^4$ interaction is non-local and non-translationally-invariant in the coordinate $\hat{\phi}$, but local within the volume of the extended object.
\subsection{Condensate: Leading order}
\begin{figure}
    \centering
    \includegraphics[width=0.98\linewidth]{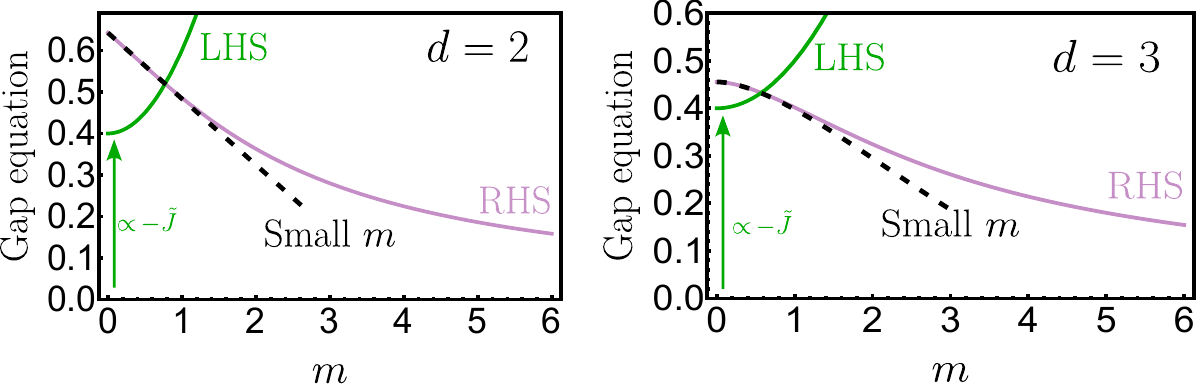}
    \caption{Terms in the gap equation Eq.~(\ref{eqn:selfconsistencym}) for $d=2$ (a) and $d=3$ (b). Right hand side (RHS) of Eq.~(\ref{eqn:selfconsistencym}) is shown in purple and has universal non analyticities as $m\to 0$ (small $m$ approximation is shown in dashed black). Left hand side (LHS) is shown in green. The height of the parabolic profile is controlled by $-\tilde{J}$ and its curvature is controlled by $\lambda$ (curvature increases as $\lambda\to 0$). Solutions relevant to field theory are those for which $m$ is small. When $\lambda<0$, the parabola opens downward.}
    \label{fig:ONGapEquation}
\end{figure}

We assume that the $N$ extended objects Bose condense and occupy only one $sp-$state $\kett{\hat{\Psi}}\approx\ket{F}\hat{F}$. Plugging this into $\hat{H}^{O(N)}$ and optimizing over $\ket{F}$ leads to the single species eigenvalue problem
\begin{equation}\label{eqn:ONSelfConsistencyWavefunction}
    \hat{h}^{O(N)}\ket{F}=\mu^{O(N)}\ket{F},
\end{equation}
where
\begin{equationS}\label{eqn:ONspHamiltonian}
    \hat{h}^{O(N)}=\frac{1}{2}\sum_{r} \hat{p}_r^2+\frac{1}{2}\sum_{r,\delta r} J_{\delta r}\hat{\phi}_r\hat{\phi}_{r+\delta r}+2\lambda \sum_{r} \sigma_r \hat{\phi}_r^2,
\end{equationS}
the Lagrange multiplier $\mu^{O(N)}$ guarantees unit normalization for $\ket{F}$, and
\begin{equation}\label{eqn:Self-Consistency}
    \sigma_r=\braket{F|\hat{\phi}_r^2|F}
\end{equation}
must be determined self-consistently. At LO, the energy of the associated quantum state is
\begin{equation}
    \hat{H}^{O(N)}=\hat{N}\bigg[\mu^{O(N)}-\lambda\sum_{r}\sigma_r^2\bigg]
\end{equation} For simplicity, we will work with translationally invariant solutions, where $\sigma_r=\sigma$ is homogeneous. Then $\hat{h}^{O(N)}$ can be diagonalized by going to a plane wave basis, with momenta denoted by $k$: $\hat{\phi}_k=V^{-1/2}\sum_ r \hat{\phi}_r e^{-i k r}$ and $\hat{p}_k=V^{-1/2}\sum_ r \hat{p}_r e^{-i k r}$. This leads to
\begin{equation}\label{eqn:GNSelfConsistent}
    \hat{h}^{O(N)}=\frac{1}{2}\sum_{k}\hat{p}_{k}^\dagger\hat{p}_k^{\vpd}+\frac{1}{2}\sum_{k}\underbrace{\Big[J_0+4\lambda\sigma-2J_2\sum_{\delta r}\cos(k_{\delta r})\Big]}_{\omega_k^2}\hat{\phi}_{k}^\dagger\hat{\phi}_k^{\vpd},
\end{equation}
with commutators $[\hat{\phi}_k,\hat{p}_{k'}]=i\delta_{k+k'}$ (the $\hat{\phi}_k$ are complex, but satisfy $\hat{\phi}^\dagger_k=\hat{\phi}_{-k}$ because of the hermiticity of $\hat{\phi}_r$, and similar relations hold for $\hat{p}_k$). These are just a bunch of harmonic oscillators for each $k$, with frequencies $\omega_k$
\begin{equation}
    \omega_k^2=(J_0+4\lambda\sigma-2J_2 d)+4J_2\sum_{\delta r}\sin(k_{\delta r}/2)^2
    \end{equation}
The $sp-$Hamiltonian can thus be rewritten as
\begin{equation}\label{eqn:ONfHamiltonian}
    \hat{h}^{O(N)}=\sum_k \omega_k\left(\hat{a}_k^\dagger\hat{a}_k+\frac{1}{2}\right),
\end{equation}
where $\hat{a}_k=\Big(\omega_k\hat{\phi}_k+i\hat{p}_k\Big)/\sqrt{2\omega_k}$ are standard annihilation operators. From this we see that, if we choose $\ket{F}$ to be the $sp-$vacuum of all $\hat{a}_k$ (i.e. $\hat{a}_k\ket{F}=0$ for all $k$), then $\mu^{O(N)}=\sum_k\omega_k/2$ is equal to
\begin{equation}
    \mu^{O(N)}=\frac{\sqrt{J_2} V}{2}\int \frac{d^dk}{(2\pi)^{d}}\Big[4\sum_{\delta r}\sin(k_{\delta r}/2)^2+m^2\Big]^{1/2},
\end{equation}
where we've introduced 
\begin{equation}
    m^2=\frac{4\lambda \sigma}{J_2} +\underbrace{\frac{J_0-2d J_2}{J_2}}_{\tilde{J}}.
\end{equation}
and defined $\tilde{J}$ as the control parameter that can be tuned to change the value of $m$. The mass parameter $m$ controls the continuum limit, defined by $m\ll  (\text{lattice spacing})^{-1}=1$. The ground state energy is then
\begin{equation}\label{eqn:ONGroundstateEnergyFull}
    \hat{H}^{O(N)}=\hat{N} V\sqrt{J_2}\bigg[\frac{1}{2}\int \frac{d^dk}{(2\pi)^d}\Big[4\sum_{\delta r}\sin(k_{\delta r}/2)^2+m^2\Big]^{1/2}-\frac{J_2^{3/2}}{16\lambda}\Big(m^2-\tilde{J}\Big)^2\bigg],
\end{equation}
while the self-consistency condition Eq.~(\ref{eqn:Self-Consistency}) becomes the well-known gap equation for $m$ (shown in Fig.~\ref{fig:ONGapEquation} for $d=2,3$):
\begin{equationS}\label{eqn:selfconsistencym}
    \sigma&=\frac{1}{V}\sum_k\braket{F|\hat{\phi}_k^\dagger\hat{\phi}_k^{\vphantom{\dagger}}|F}\\[5pt]
    \rightarrow\frac{J_2}{4\lambda}\Big(m^2-\tilde{J}\Big)&=\int \frac{d^d k}{2\sqrt{J_2}(2\pi)^d}\Big[4\sum_{\delta r}\sin(k_{\delta r}/2)^2+m^2\Big]^{-1/2}.
\end{equationS}
The massless point (i.e. existence of a solution with $m=0$) is obtained by tuning $\tilde{J}$ adequately and only exists for $d\geq 2$. In various dimensions, small $m$ expressions for the Hamiltonian (setting $J_2=1$ for simplicity) are
\begin{equationS}
    % \hat{H}^{O(N)}_{d=1}&=\hat{N}V\bigg[\frac{2}{\pi}-\frac{m^2}{8\pi}\log\bigg(\frac{m^2}{64e}\bigg)-\frac{1}{16\lambda}\Big(m^2-\tilde{J}\Big)^2+O(m^4\log m)\bigg]\\[8pt]
    \hat{H}^{O(N)}_{d=2}&=\hat{N}V\bigg[A_2+B_2 m^2-\frac{|m|^3}{12\pi}+C_2m^4-\frac{1}{16\lambda}\Big(m^2-\tilde{J}\Big)^2+O(m^6)\bigg]\\[8pt]
    \hat{H}^{O(N)}_{d=3}&=\hat{N}V\bigg[A_3+B_3 m^2+\frac{m^4}{64\pi^2}\log m^2+C_3m^4-\frac{1}{16\lambda}\Big(m^2-\tilde{J}\Big)^2+O(m^6\log m)\bigg]
\end{equationS}
We give expressions for the numerical prefactors $A_2,A_3,B_2,B_3,C_2,C_3$ in Appendix~\ref{app:ONNumCoeff}, but they are of order $\sim 1$ and depend on the microscopic details of the model. As is well known, the small $m$ expansion generates universal terms $\propto m^3,m^4\log m$ whose coefficients are independent of the microscopic model~\cite{MOSHE2003}. The self-consistency gap equation in various dimensions for small $m$ is
% \begin{align}\begin{split}
%     % \frac{1}{2\lambda}\Big(m^2-\tilde{J}\Big)_{d=1}&=-\frac{1}{2\pi}\log\bigg(\frac{m^2}{64}\bigg)+O(m^2\log m)\rightarrow m^2\approx 64 \exp\bigg(\frac{\pi \tilde{J}}{\lambda}\bigg)\\[8pt]
%     \frac{1}{2\lambda}\Big(m^2-\tilde{J}\Big)_{d=2}&=-\frac{|m|}{2\pi}+0.64+O(m^2)\rightarrow |m|_*\approx \frac{2\pi \Big(0.64+\frac{\tilde{J}}{2\lambda}\Big)}{\frac{1}{2}+\sqrt{\frac{1}{4}+\Big(\frac{2\pi^2}{\lambda}\Big)\Big(0.64+\frac{\tilde{J}}{2\lambda}\Big)}}\\[8pt]
%     \frac{1}{2\lambda}\Big(m^2-\tilde{J}\Big)_{d=3}&=0.46+\frac{m^2}{8\pi^2}\log(0.021m^2)+O(m^4\log m)\rightarrow m^2_*\approx\frac{8\pi^2\Big(\frac{\tilde{J}}{2\lambda}+0.46\Big)}{\frac{4\pi^2}{\lambda}-\log \Big[1.66\Big(\frac{\tilde{J}}{2\lambda}+0.46\Big)\Big]}.
% \end{split}\end{align}
\begin{align}\begin{split}\label{eqn:selfgeneratedmass}
    \frac{1}{2\lambda}\Big(m^2-\tilde{J}\Big)_{d=2}&=-\frac{|m|}{2\pi}+4B_2+8C_2m^2+O(m^4)\\[8pt]
    \frac{1}{2\lambda}\Big(m^2-\tilde{J}\Big)_{d=3}&=4B_3+\frac{m^2}{8\pi^2}\log m^2+m^2\left(\frac{1}{16\pi^2}+8C_3\right)+O(m^4\log m)
\end{split}\end{align}
\subsubsection{Solutions to the gap equation}
If $\lambda\sim 1$ in $d=2$, we can neglect the $m^2$ terms in Eq.~(\ref{eqn:selfgeneratedmass}) to find that
\begin{equationS}
    m'&=\frac{\pi}{\lambda}(8\lambda B_{2}+\tilde{J})\\
    \hat{H}^{O(N)}_{d=2}&=(A_2-4\lambda B_2^2)+m'\left(\frac{B_2\lambda}{\pi}\right)-m'^2\left(\frac{\lambda}{16\pi^2}\right)+\frac{m'^3}{24\pi},
\end{equationS}
where we have replaced $\tilde{J}$ in favor of $m'$, and $\tilde{J}$ needs to be tuned to make $m'$ small, which is required for self-consistency. Since $\lambda$ has dimensions of energy in $d=2$, the condition $\lambda\sim 1$ means that $\lambda$ is at the lattice scale.

If $\tilde{J}$ is tuned to $\tilde{J}_c=-8\lambda B_{2,3}$, then $m=0$ is exactly a solution for both $d=2,3$. At this precise value of $\tilde{J}$ the gap equation posseses another solution, with a dynamically generated mass:
\begin{equationS}\label{eqn:ONdefm}
    |m|_{d=2}&\equiv\tilde{m}_{d=2}=-\frac{\lambda}{\pi(1-16\lambda C_2)}\\
    m_{d=3}&\equiv \tilde{m}_{d=3}=\exp\left(\frac{2\pi^2}{\lambda}-\frac{1}{4}-32\pi^2 C_3\right),
\end{equationS}
where we've defined $\tilde{m}$ for future reference. In both cases $\lambda$ needs to be negative and small so that $\tilde{m}\ll 1$ (although $\lambda=-1$ is enough for $d=3$ due to the exponential sensitivity), which is the assumption under which these solutions were found, and is required for a well-defined continuum limit. We omit here considerations of stability for $\lambda<0$, which have been heavily discussed in the literature~\cite{Moshe1983,Giombi2020} and defer them to Appendix~\ref{app:ONMetastable}, where we analyze a more general potential $U(\phi^2)$, following Ref.~\cite{MOSHE2003}. We will just mention that, within this lattice construction, the associated solutions are metastable with respect to other minima of $U$ that do not have $m\ll 1$ and thus do not represent continuum solutions. In any case, these values of $\tilde{m}$ lead to the following ground state energies
\begin{equationS}
    \hat{H}^{O(N)}_{d=2}&\approx\hat{N}V\bigg[\left(A_2-\frac{\tilde{J}^2}{16\lambda}\right)-\frac{\tilde{m}^3}{48\pi}\bigg]\\[8pt]
    \hat{H}^{O(N)}_{d=3}&\approx\hat{N}V\bigg[\left(A_3-\frac{\tilde{J}^2}{16\lambda}\right)-\frac{\tilde{m}^4}{128\pi^2}\bigg].
\end{equationS}
In comparison, the energies for the $m=0$ solutions only include the terms in parentheses, so $m=\tilde{m}$ is the preferred vacuum. Such non-perturbatively generated mass scales arise in other systems, and have been conjectured to provide an alternate mechanism for gauge boson mass generation in the electroweak sector of the Standard Model~\cite{romatschke2024mass}. If $\lambda>0$, then $m=0$ is the only solution. 

\subsubsection{Correlators}\label{subsubsec:LOCorrelators}
In the Hamiltonian formalism it is very natural to consider equal-time correlators. To begin with, we have operators like
\begin{equation}\label{eqn:LOCorrelatorSymmetryPreservingPhase}
    \sum_{\alpha}\hat{\phi}_{r}^{\alpha}\hat{\phi}_{r+R}^\alpha=\bbrakettPsi{\hat{\phi}_r\hat{\phi}_{r+R}}
\end{equation}
The leading order contribution to this operator will be a classical number times $\hat{N}$, given by  
\begin{equationS}
    \bbrakettPsi{\hat{\phi}_r\hat{\phi}_{r+R}}&=\frac{\hat{N}}{V}\sum_r\braket{F|\hat{\phi}_r\hat{\phi}_{r+R}|F}+O(\hat{N}^{1/2})=\frac{\hat{N}}{V}\sum_k\frac{e^{-ik\cdot R}}{2\omega_k}
\end{equationS}
with correlations characteristic of free fields
% \begin{equationS}
%     \frac{1}{V}\sum_{r}\bbrakettPsi{\hat{\phi}_r\hat{\phi}_{r+R}}_{d=2}&=\hat{N}\begin{dcases}
%         \frac{1}{4\pi |R|} & 1\ll |R|\ll m^{-1} \\
%         \frac{e^{-m|R|}}{4\pi |R|} & 1\ll m^{-1}\ll |R|
%     \end{dcases}\\[10pt]
%     \frac{1}{V}\sum_{r}\bbrakettPsi{\hat{\phi}_r\hat{\phi}_{r+R}}_{d=3}&=\hat{N}\begin{dcases}
%         \frac{1}{\pi^2 |R|^2} & 1\ll |R|\ll m^{-1} \\
%         \sqrt{\frac{m}{32\pi}}\frac{e^{-m|R|}}{R^{3/2}} & 1\ll m^{-1}\ll |R|
%     \end{dcases}
% \end{equationS}
\begin{equationS}
    \bbrakettPsi{\hat{\phi}_r\hat{\phi}_{r+R}}_{d=2}&\sim\hat{N}\begin{dcases}
       |R|^{-1}& 1\ll |R|\ll m^{-1} \\
        |R|^{-1} e^{-m|R|}& 1\ll m^{-1}\ll |R|
    \end{dcases}\\[10pt]
    \bbrakettPsi{\hat{\phi}_r\hat{\phi}_{r+R}}_{d=3}&\sim\hat{N}\begin{dcases}
        |R|^{-2} & 1\ll |R|\ll m^{-1} \\
        |R|^{-3/2}e^{-m|R|} & 1\ll m^{-1}\ll |R|
    \end{dcases}.
\end{equationS}
Products of ``single trace" operators factorize here too (to leading order):
\begin{equation}
  \left(\sum_{\alpha}\hat{\phi}_{r}^{\alpha}\hat{\phi}_{r}^\alpha\right)\left(\sum_{\alpha}\hat{\phi}_{r+R}^{\alpha}\hat{\phi}_{r+R}^\alpha\right)= \hat{N}^2\braket{F|\hat{\phi}_r^2|F}\braket{F|\hat{\phi}_{r+R}^2|F}+O(\hat{N}^{3/2}).
\end{equation}
% These solutions can be used to get rid of $\tilde{J}$ and express everything in terms of $m$ and $J_2^{3/2}/\lambda$:
% \begin{equationS}
%     \hat{H}^{O(N)}_{d=1}&=\hat{N}V\sqrt{J_2}\bigg[\frac{2}{\pi}+\frac{m^2}{8\pi}-\bigg(\frac{\lambda}{16\pi^2 J_2^{3/2}}\bigg)\log\bigg(\frac{m^2}{64}\bigg)^2+\frac{J_2^{3/2}m^4}{16\lambda}+O(m^4\log m)\bigg]\\[8pt]
%     \hat{H}^{O(N)}_{d=2}&=\hat{N}V\sqrt{J_2}\bigg[0.48+0.16 m^2+\frac{|m|^3}{24\pi}-\frac{J_2^{3/2}\tilde{J}^2}{16\lambda}+m^2\bigg(\frac{J_2^{3/2} \tilde{J}}{8\lambda}\bigg)+O(m^4)\bigg]
% \end{equationS}

\subsubsection{Symmetry-broken phase}
 In $d=2,3$ the gap equation in Eq.~(\ref{eqn:selfconsistencym}) ceases to have real solutions when $\tilde{J}<\tilde{J}_c$, which signals the onset of a symmetry-breaking transition. This process and the symmetry-broken phase can also be described using the fluctuating $N$ formalism. At LO, where we are concerned only with the $sp-$state $ \ket{F}$, we see from Eq.~(\ref{eqn:GNSelfConsistent}) that, as $m\to 0$, then $\braket{F|\hat{\phi}_{k=0}^\dagger\hat{\phi}_{k=0}|F}=1/(2m)\equiv \upsilon$ starts growing without bounds and becomes extensive, so we have to treat it separately from the rest of $k$ modes when calculating $\sigma_r$ using Eq.~(\ref{eqn:Self-Consistency}). With this refinement, the gap equation reads:
 \begin{equationS}
     \sigma&=\frac{1}{V}\braket{F|\hat{\phi}_0^\dagger\hat{\phi}_0^{\vphantom{\dagger}}|F}+\frac{1}{V}\sum_{k\neq 0}\braket{F|\hat{\phi}_k^\dagger\hat{\phi}_k^{\vphantom{\dagger}}|F}\\[5pt]
     \rightarrow\frac{1}{4\lambda}(-\tilde{J})&=\frac{\upsilon}{V}+\int \frac{d^dk}{2(2\pi)^d\left[4\sum_{\delta r}\sin(k_{\delta r}/2)^2\right]^{1/2}}
 \end{equationS}
where we have set $m=0$ everywhere except in $\upsilon$. Because of this the momentum space integral is just a number, and the gap equation becomes  a relation between $\tilde{J}$ and $\upsilon$
\begin{equation}
   \frac{\tilde{J}_c-\tilde{J}}{4\lambda}=\frac{\upsilon}{V}\equiv\rho_0.
\end{equation}
This further shows that $m\sim 1/V$. Formulas for all observables must take into account the macroscopic size of the $k=0$ mode. For instance [compare with Eq.~(\ref{eqn:LOCorrelatorSymmetryPreservingPhase})],
 \begin{equation}
     \bbrakettPsi{\hat{\phi}_r\hat{\phi}_{r+R}}\approx\frac{\hat{N}}{V}\sum_k\braket{F|\hat{\phi}_k^\dagger\hat{\phi}_k|F}e^{-ikR}=\hat{N}\rho_0+\hat{N}\int\frac{d^dk}{(2\pi)^d}\frac{e^{-ik R}}{2\omega_k},
 \end{equation}
 where $\omega_k$ here is evaluated at $m=0$.

\subsection{Excitations: Next to leading order}
To obtain the excitation spectrum in the symmetry preserving phase, we proceed similarly to the case of the quantum mechanical model. We extract the condensate operators from $\kett{\hat{\Psi}}=\hat{F}\ket{F}+\kett{\delta\hat{\Psi}}$, we represent the occupation of the condensate ($\hat{F}^\dagger\hat{F}$) in terms of $\hat{N}$ and the occupation of the excitations using $\hat{F}^\dagger\hat{F}=\hat{N}-\bbrakett{\delta\hat{\Psi}|\delta\hat{\Psi}}$, and collect terms according to their power of $\hat{N}$. To order $\sim N^0$, the Hamiltonian is
\begin{equationS}
    \hat{H}^{O(N)}_{\text{NLO}}&=\hat{N}V \epsilon+\bbrakett{\delta\hat{\Psi}|\left(\hat{h}_*^{O(N)}-\mu^{O(N)}\right)|\delta\hat{\Psi}}+\lambda\sum_{r}\Big(\hat{\Phi}_r+\hat{\Phi}^\dagger_r\Big)^2,
\end{equationS}
where $\epsilon$ is the energy density per species obtained in the leading order section [Eq.~(\ref{eqn:ONGroundstateEnergyFull})],
\begin{equation}\label{eqn:PhiR}
    \hat{\Phi}_r=\brakett{F|\hat{\phi}_r^2|\delta\hat{\Psi}} e^{-i\hat{\varphi}}
\end{equation}
is an operator in the multispecies Hilbert space and $\hat{h}^{O(N)}_*$ is the $sp-$Hamiltonian from Eq.~(\ref{eqn:GNSelfConsistent}) evaluated at $m=m_*$, where $m_*$ is any of the solutions to the gap equation. As in the previous section, we expand $\kett{\delta\hat{\Psi}}$ in terms of the $sp-$eigenstates of $\hat{h}_*^{O(N)}$, which we denote $\ket{n}$ (with associated eigenvalue $\varepsilon_n$), so that 
\begin{equationS}
    \hat{H}^{O(N)}_{\text{NLO}}&=\hat{N} V\epsilon +\sum_n (\varepsilon_n-\mu^{O(N)})\hat{\Psi}^\dagger_n\hat{\Psi}_n+\lambda\sum_{r}\big(\hat{\Phi}_r+\hat{\Phi}^\dagger_r\big)^2.
\end{equationS}
Since $\ket{n}$ is an $sp-$eigenstate of $\hat{h}^{O(N)}_*$, each different $n$ corresponds to a possible configuration of occupation numbers of the $k$-modes in Eq.~(\ref{eqn:ONfHamiltonian}). Thus, the system has an extensive number of excitation \textit{types}, each of which can in turn be excited to a different level.
\subsubsection{Interacting sector}
The $sp-$eigenstates of $\hat{h}^{O(N)}_*$ (and hence the operators $\hat{\Psi}_n$ in $\kett{\delta\hat{\Psi}}=\sum_{n\neq F}\ket{n}\hat{\Psi}_n$) are labeled by occupation numbers of an infinity of momentum modes, the number of which is unwieldy. Nevertheless, most $\hat{\Psi}_n$ won't appear directly in $\hat{\Phi}_r$ and hence will only contribute trivially to the Hamiltonian. We thus need to investigate which operators $\hat{\Psi}_n$ appear in $\hat{\Phi}_r$. To do this, we expand $\hat{\phi}_r$ in Eq.~(\ref{eqn:PhiR}) in terms of the $sp-$annihilation operator $\hat{a}_k$, noting that $\ket{F}$ is the $sp-$vacuum of all the $\hat{a}_k$. This leads to 
\begin{equation}
    \hat{\Phi}_r=\frac{1}{2V}\sum_{kk'}\frac{e^{i(k+k')r}}{\sqrt{\omega_{k}\omega_{k'}}}\brakett{F|\hat{a}_{k}\hat{a}_{k'}|\delta\hat{\Psi}}e^{-i\hat{\varphi}}
\end{equation}
The $sp-$state $\hat{a}_{k}^\dagger\hat{a}_{k'}^\dagger\ket{F}$ is an $sp-$eigenstate of $\hat{h}^{O(N)}_*$, with eigenvalue $\mu^{O(N)}+\omega_{k}+\omega_{k'}$, and thus one of the labels $n$ in $\hat{\Psi}_{n}$ can be chosen to correspond precisely to this $sp-$state. We thus define the object
\begin{equation}
    \hat{\Psi}_{kk'}^{\vpd}=\brakett{F|\hat{a}_{k}\hat{a}_{k'}|\delta\hat{\Psi}}e^{-i\hat{\varphi}},
\end{equation}
which is symmetric in $k\leftrightarrow k'$, and satisfies
\begin{equation}
    \big[\hat{\Psi}_{kk'}^{\vphantom{\dagger}},\hat{\Psi}^\dagger_{qq'}\big]=\braket{F|\hat{a}_k^{\vpd}\hat{a}_{k'}^{\vpd}\hat{a}^\dagger_q\hat{a}^\dagger_{q'}|F}=\delta_{kq}\delta_{k'q'}+\delta_{kq'}\delta_{k'q}.
\end{equation}
Explicitly keeping track of these operators, the Hamiltonian becomes
\begin{equationS}\label{eqn:NLOHamiltoniainONModel}
    \hat{H}^{O(N)}_{\text{NLO}}&=\hat{N}V\epsilon+\sum_n' (\varepsilon_n-\mu^{O(N)})\hat{\Psi}^\dagger_n\hat{\Psi}_n+\frac{1}{2}\sum_{kk'}(\omega_{k}+\omega_{k'})\hat{\Psi}^\dagger_{kk'}\hat{\Psi}_{kk'}^{\vpd}\\
    &+\frac{\lambda}{4V}\sum_{k k' q q'}\frac{\delta_{k+k',q+q'}}{\sqrt{\omega_{k}\omega_{k'}\omega_{q}\omega_{q'}}}\big(\hat{\Psi}^{\vpd}_{kk'}+\hat{\Psi}^\dagger_{-k,-k'}\big)\big(\hat{\Psi}^{\dagger}_{qq'}+\hat{\Psi}^{\vpd}_{-q,-q'}\big),
\end{equationS}
where the prime in the sum over $n$ indicates that the modes $\hat{\Psi}_{kk'}$ must not be included, the factor of $1/2$ in front of the $kk'$ sum accounts for double counting, and the Kronecker delta in the $kk'qq'$ sum enforces momentum conservation, i.e. only configurations with $k+k'=q+q'$ contribute. This Hamiltonian is quadratic and can be diagonalized efficiently. Nevertheless, it still describes scattering in pairs, whereby a pair $kk'$ can be scattered into a pair $qq'$ with some amplitude, subject only to momentum conservation. 
\subsubsection{O(N) invariance}
As in the quantum mechanical case, the excitations described by the previous Hamiltonian are all permutationally symmetric, but not necessarily $O(N)$ invariant. The $O(N)$ invariance constraint leads, as in Section~\ref{sec:QMONInvariance}. to an infinity of conditions that $O(N)$ invariant states $\ket{S}$ must satisfy (see Appendix~\ref{app:ONConstraint}),
\begin{equation}\label{eqn:ONConstraint}
    \left[\sum_k (\hat{a}_k^\dagger\hat{b}^{\vpd}_{k}-\hat{b}^\dagger_k\hat{a}^{\vpd}_k)\kett{\hat{\Psi}}_a\otimes\kett{\hat{\Psi}}_b\right]\ket{S}=0
\end{equation}
When $N\to\infty$ (at fixed $V$), this leads to
\begin{equation}\label{eqn:ONconstraintLeadingOrder}
    \hat{\Psi}_n\ket{S}=0,\text{ for }n\neq kk',
\end{equation}
and only the modes $\hat{\Psi}_{kk'}$ can be excited. The $O(N)$ invariant spectrum of the system is thus described entirely by
\begin{equationS}
    \hat{H}^{O(N)}_{\text{NLO}}=\hat{N}V\epsilon&+\frac{1}{2}\sum_{kk'}(\omega_{k}+\omega_{k'})\hat{\Psi}^\dagger_{kk'}\hat{\Psi}_{kk'}^{\vpd}\\
    &+\frac{\lambda}{4V}\sum_{k k' q q'}\frac{\delta_{k+k',q+q'}}{\sqrt{\omega_{k}\omega_{k'}\omega_{q}\omega_{q'}}}\big(\hat{\Psi}^{\vpd}_{kk'}+\hat{\Psi}^\dagger_{-k,-k'}\big)\big(\hat{\Psi}^{\dagger}_{qq'}+\hat{\Psi}^{\vpd}_{-q,-q'}\big).
\end{equationS}
$O(N)$ invariance naturally leads to a description in terms of fields with two $k$ indices, which are the Fourier transforms of bilocal fields with two position indices. %In the multiparticle interpretation, the operator $\hat{\Psi}_{kk'}$ creates an extended object in the $sp-$state $\hat{a}^\dagger_k\hat{a}^\dagger_{k'}\ket{F}$ which is a $sp-$state with two $k$ excitations.
Correspondingly, the other excitations of the system, which correspond to other occupations of $sp-$states and are not $O(N)$ invariant, have more $k$ indices and a multilocal structure.
\subsubsection{Bound states at negative coupling constant}\label{subsubsec:ONBoundStates}
When $\lambda$ is negative the system is known to posses bound states~\cite{Abbott1976,ROMATSCHKE2023138270}. In the Hamiltonian formalism, these can be calculated by diagonalizing $\hat{H}^{O(N)}_{\text{NLO}}$, which is formally a system of linearly coupled harmonic oscillators. The normal modes of this ensemble of oscillators will have associated excitation frequencies, the lowest of which corresponds to the bound state mass. 

To proceed, we switch the labels of $\hat{\Psi}_{kk'}$ from $k,k'$ to total momentum $P=k+k'$ and momentum transfer $Q=k-k'$. We do the same to $\hat{\Psi}_{qq'}$, with $\tilde{P}=q+q'$ and $\tilde{Q}=q-q'$. Noting that different $P$ decouple because of momentum conservation, we have
\begin{equationS}
    \hat{H}^{O(N)}_{\text{NLO}}=\hat{N}V\epsilon&+\frac{1}{2}\sum_{P,Q}(\omega_{k}+\omega_{k'})\hat{\Psi}^\dagger_{QP}\hat{\Psi}_{QP}^{\vpd}\\
    &+\frac{\lambda}{4V}\sum_{P,Q,\tilde{Q}}\frac{\big(\hat{\Psi}^{\vpd}_{PQ}+\hat{\Psi}^\dagger_{-P,-Q}\big)\big(\hat{\Psi}^{\dagger}_{P,\tilde{Q}}+\hat{\Psi}^{\vpd}_{-P,-\tilde{Q}}\big)}{\sqrt{\omega_{k}\omega_{k'}\omega_{q}\omega_{q'}}}.
\end{equationS}
We have kept $k,k',q,q'$ in the expressions, but they are now implicitly defined as $k=(P+Q)/2$, $k'=(P-Q)/2$, $q=(P+\tilde{Q})/2$ and $q'=(P-\tilde{Q})/2$. With this relabeling, we have that $\hat{\Psi}_{PQ}=\hat{\Psi}_{P,-Q}$ because of $k\leftrightarrow k'$ symmetry, and
\begin{equation}
    \left[\hat{\Psi}_{PQ}^{\vpd},\hat{\Psi}_{\tilde{P},\tilde{Q}}^\dagger\right]=\delta_{P\tilde{P}}\left(\delta_{Q\tilde{Q}}+\delta_{Q,-\tilde{Q}}\right).
\end{equation}
To avoid double counting, we will restrict sums in $Q$ to half their allowed values. Note also that, while all operators with $Q\neq 0$ have standard commutators, the $Q=0$ modes have a non-standard normalization. We now introduce quadrature operators
\begin{equationS}
    \hat{X}_{PQ}&=\frac{\hat{\Psi}_{PQ}^{\vpd}+\hat{\Psi}_{-P,Q}^\dagger}{\sqrt{2(1+\delta_{Q0})}}\\
    \hat{\Pi}_{PQ}&=\frac{\hat{\Psi}_{PQ}^{\vpd}-\hat{\Psi}_{-P,Q}^\dagger}{i\sqrt{2(1+\delta_{Q0})}},
\end{equationS}
with commutation relations (in the restricted set of $Q$)
\begin{equation}
    \left[\hat{X}_{PQ},\hat{\Pi}_{\tilde{P}\tilde{Q}}\right]=i\delta_{P,-\tilde{P}}\,\delta_{Q\tilde{Q}},
\end{equation}
to re-express the Hamiltonian as
\begin{equationS}
    \hat{H}^{O(N)}_{\text{NLO}}&=\hat{N}V\epsilon-\frac{1}{2}\sum_{P,Q}'(\omega_k+\omega_{k'})+\sum_{P,Q}'(\omega_{k}+\omega_{k'})\left(\frac{\hat{\Pi}_{P,Q}^\dagger\hat{\Pi}_{P,Q}^{\vpd}}{2}\right)
    \\&+\sum_{P,Q,\tilde{Q}}'\frac{\hat{X}_{P,\tilde{Q}}^\dagger\hat{X}_{P,Q\vphantom{\tilde{Q}}}^{\vpd}}{2}\left[(\omega_k+\omega_{k'})\delta_{Q\tilde{Q}}+\frac{4\lambda}{V}\frac{\eta_Q^{\vpd}\eta_{\tilde{Q}}}{\sqrt{\omega_{k}\omega_{k'}\omega_{q}\omega_{q'}}}\right],
\end{equationS}
where the prime in the sum indicates that only the restricted set of $Q$'s should be considered and $\eta_Q=1/\sqrt{2}$ if $Q=0$ and $1$ otherwise. We further define $\Theta_{P,Q}=\sqrt{
\omega_{k}+\omega_{k'}}\, \Pi_{P,Q}$ and $Y_{P,Q}= X_{P,Q}/\sqrt{
\omega_{k}+\omega_{k'}}$, to make the $\hat{\Theta}_{P,Q}^\dagger\hat{\Theta}_{P,Q}^{\vpd}$ terms have identical prefactors:
\begin{equationS}
    \hat{H}^{O(N)}_{\text{NLO}}&=\hat{N}V\epsilon-\frac{1}{2}\sum_{P,Q}'(\omega_k+\omega_{k'})+\sum_{P,Q}'\left(\frac{\hat{\Theta}_{P,Q}^\dagger\hat{\Theta}_{P,Q}^{\vpd}}{2}\right)
    \\&+\sum_{P,Q,\tilde{Q}}'\frac{\hat{Y}_{P,\tilde{Q}}^\dagger\hat{Y}_{P,Q\vphantom{\tilde{Q}}}^{\vpd}}{2}\underbrace{\left[(\omega_k+\omega_{k'})^2\delta_{Q\tilde{Q}}+\frac{4\lambda}{V}\frac{\eta_Q^{\vpd}\eta_{\tilde{Q}}\sqrt{(\omega_k+\omega_{k'})(\omega_q+\omega_{q'})}}{\sqrt{\omega_{k}\omega_{k'}\omega_{q}\omega_{q'}}}\right]}_{L^P_{Q\tilde{Q}}},
\end{equationS}
The excitation frequencies of the normal modes of the system can be determined from the coefficients in front of the $\hat{Y}_{P,\tilde{Q}}\hat{Y}_{P,Q\vphantom{\tilde{Q}}}$ terms
\begin{equation}
    L_{Q,\tilde{Q}}^P=\left[(\omega_k+\omega_{k'})^2\delta_{Q\tilde{Q}}+\frac{4\eta_Q^{\vpd}\eta_{\tilde{Q}}\lambda}{V}\frac{\sqrt{(\omega_k+\omega_{k'})(\omega_q+\omega_{q'})}}{\sqrt{\omega_{k}\omega_{k'}\omega_{q}\omega_{q'}}}\right],
\end{equation}
which should be thought of as a matrix $L^P$ with indices $Q,\tilde{Q}$, and where $P$ is just a parameter. The excitation frequencies are then the square roots of the eigenvalues of $L$, which we denote as $l$, and are obtained from the equation~\cite{Maslov199} (see also Appendix~\ref{app:ONBoundState})
\begin{equation}\label{eqn:BoundStateEquation}
    -1=\frac{4\lambda}{V}\sum_Q'\eta_Q^2\left(\frac{1}{\omega_k}+\frac{1}{\omega_{k'}}\right)\left[\frac{1}{(\omega_k+\omega_{k'})^2-l}\right]\equiv B(l,P).
\end{equation}
For the bound state, we have that $E_{\text{bound}}^2=l<\min_{Q}(\omega_{k}+\omega_{k'})^2$, so the denominator inside the $Q$ sum has no singularities and we can naively take the infinite volume limit. If we tune $\tilde{J}$ so that $m=\tilde{m}_{d=2,3}$ (\ref{eqn:ONdefm}, i.e. the configuration with a dynamically generated mass), and let  $m, |P|\ll 1$, we find that $\theta\equiv\sqrt{(E_{\text{bound}}^2-P^2)/(2m)^2}$ satisfies (see Appendix~\ref{app:ONBoundState})
\begin{equationS}
    1=\frac{1}{4\theta}\log\left(\frac{1+\theta}{1-\theta}\right)\rightarrow \theta=0.956\rightarrow \boxed{E_{\text{bound}}^2=P^2+(1.92\tilde{m})^2}
\end{equationS}
in $d=2$, and
\begin{equationS}
    \frac{1}{2}=\left(\sqrt{\frac{1}{\theta^2}-1}\right)\arctan\left(\frac{1}{\sqrt{\frac{1}{\theta^2}-1}}\right)\rightarrow \theta=0.919\rightarrow \boxed{E_{\text{bound}}^2=P^2+(1.84\tilde{m})^2},
\end{equationS}
in $d=3$, in agreement with known results~\cite{Abbott1976,ROMATSCHKE2023138270}. Other eigenvalues of $L_P$ give rise to the energies of scattering states.
\subsubsection{Ground state energy}
The ground state energy up to NLO can be expressed in terms of the matrix $L^P$ as
\begin{equation}
    E^{O(N)}_{\text{GND}}=\hat{N}V\epsilon+\frac{1}{2}\sum_P\mathrm{Tr}\left(\sqrt{L^P}\right)-\frac{1}{2}\sum_{P,Q}'(\omega_k+\omega_{k'})
\end{equation}
and represents the change in zero-point energy of the oscillators ($\sum_P\mathrm{Tr}(\sqrt{L^P})$ is the sum over the normal mode frequencies). To calculate this, we introduce the trace of the resolvent
\begin{equation}
    R(z)=\sum_P\mathrm{Tr}\left(\frac{1}{L^P-z}\right),
\end{equation}
which is a function of the complex variable $z$. The eigenvalues of the matrix $L^P$ are the poles of $R(z)$ and in the infinite volume limit many of them coalesce into a branch cut of finite extent (due to the finite lattice spacing), except for the bound state, which remains a pole. Using $R(z)$ we can calculate trace of functions of $L^P$ by integration over a countour $\mathcal{C}_1$ that encircles all these non-analyticities:
\begin{equation}
    \sum_P\mathrm{Tr}\left[g(L^P)\right]=\oint_{\mathcal{C}_1}\frac{g(z)dz}{2\pi i }R(z).
\end{equation}
Given that $L$ is a diagonal matrix plus a rank one perturbation it follows that $R(z)$ can be calculated explicitly,
\begin{equation}\label{eqn:Resolventt}
    R(z)=\sum_{Q,P}'\frac{1}{(\omega_k+\omega_{k'})^2-z}-\sum_P\frac{\partial_zB(z,P)}{1+B(z,P)},
\end{equation}
where $B(z,P)$ is defined in Eq.~(\ref{eqn:BoundStateEquation}). For real $z>0$, $B(z,P)$ also has a branch cut of finite extent that begins, for small $|P|$, at $z=4m^2+P^2$ and reflects the presence of scattering states above a threshold energy. The ground state energy is then
\begin{equation}
    E^{O(N)}_{\text{GND}}=\hat{N}V\epsilon-\frac{1}{2}\sum_P\oint_{\mathcal{C}_1}\frac{\sqrt{z}\,dz}{2\pi i}\frac{\partial_zB(z,P)}{1+B(z,P)}
\end{equation}
The contour $\mathcal{C}_1$ makes manifest some physical features of the theory, like separating the contributions from the zero-point energy of the bound state and that of the scattering states. However, for comparison with other works it is better to deform the contour to the negative real axis, picking up along the way the contributions from the branch cut of the square root:
\begin{equationS}\label{eqn:ONNLOGNDEnergy}
    E^{O(N)}_{\text{GND}}&=\hat{N}V\epsilon-\frac{1}{2\pi}\sum_P\int_0^{\infty}\frac{s\,ds\,\partial_sB(-s^2,P)}{1+B(-s^2,P)}\\
    &=\hat{N}V\epsilon+\frac{1}{2\pi}\sum_P\int_0^{\infty}\,ds\log\left[1+B(-s^2,P)\right].
\end{equationS}
We have parameterized the new contour using $z=-s^2$ and integrated by parts with respect to $s$ ($B$ goes like $1/s^2$ at large $s$, so we don't pick any contribution from $s=\infty$). For $\lambda>0$ this is a safe procedure since $1+B(-s^2,P)>0$ [see Eq.~(\ref{eqn:BoundStateEquation})]. For $\lambda<0$ there might be a pole at some finite $s$ but its contribution vanishes because it contributes with opposite signs on each side of the square root branch cut, and the integral must be taken in the principal value sense. For $|P|,s,m\ll 1$ the function $B(-s^2,P)$ can be expressed as
\begin{equation}
    1+B(-s^2,P)=b(s^2+P^2),
\end{equation}
where
\begin{equationS}
   b(u^2)_{d=2}&=\frac{\lambda}{\pi}\left[-\frac{1}{\tilde{m}_{d=2}}+\frac{1}{u}\arctan\left(\frac{u}{2m}\right)\right]\\
   b(u^2)_{d=3}&=\frac{\lambda}{4\pi^2}\left[\log\left(\frac{\tilde{m}^2_{d=3}}{m^2}\right)+1-2\left(\sqrt{1+\frac{4m^2}{u^2}}\right)\mathrm{atanh}\left(\frac{u}{\sqrt{u^2+4m^2}}\right)\right],
\end{equationS}
and $\tilde{m}$ are defined in Eq.~(\ref{eqn:ONdefm}). By comparison with other works~\cite{Abbott1976,Romatschke2023} it is clear that $1+B(-s^2,P)$ is proportional to the inverse propagator of the $\hat{\phi}^2$ composite field evaluated at Euclidean momenta [note that this proportionality factor does not affect the ground state energy, see Eq.~(\ref{eqn:ONNLOGNDEnergy})]. Exact expressions for the energy will depend on the microscopic model under consideration, but the region of small $|P|, s$ will generate universal terms. In $d=2$, the energy will have the form
\begin{equation}
    \hat{H}^{O(N)}=A'+B'm'^2+V\hat{N}\frac{m'^3}{24\pi}\left[1-\frac{32}{\pi^2 \hat{N}}\left(1-\frac{4}{\pi^2}\right)\log\left(\frac{m'}{C'}\right)\right]+O(m^4),
\end{equation}
with $A',B',C'$ being model dependent. The $\log m'$ term contains a contribution from the critical exponent $\alpha= 1-32/(\pi^2 N)+O(N^{-2})$, and an additional piece expected for specific spatial dimensionalities~\cite{Abe1973,ABE197311,Aharony1974,Wegner1972}.
\subsubsection{Correlators}
The calculation of the correlators in section~\ref{subsubsec:LOCorrelators} can be extended into higher orders. Keeping all the terms in the expansion of $\kett{\hat{\Psi}}$, we have
\begin{equationS}
    \bbrakettPsi{\hat{\phi}_r\hat{\phi}_{r+R}}&=\hat{F}^\dagger\hat{F}\braket{F|\hat{\phi}_r\hat{\phi}_{r+R}|F}+\left(\hat{F}\bbraket{\delta\hat{\Psi}|\hat{\phi}_r\hat{\phi}_{r+R}|F}+\hat{F}^\dagger\brakett{F|\hat{\phi}_r\hat{\phi}_{r+R}|\delta\hat{\Psi}}\right)\\
    &+\bbrakett{\delta\hat{\Psi}|\hat{\phi}_r\hat{\phi}_{r+R}|\delta\hat{\Psi}}.
\end{equationS}
Expanding to $O(\hat{N}^{1/2})$ and replacing $\hat{\phi}_r$ by $sp-$annihilation operators yields
\begin{equationS}
    \bbrakettPsi{\hat{\phi}_r\hat{\phi}_{r+R}}&=\hat{N}\braket{F|\hat{\phi}_r\hat{\phi}_{r+R}|F}+\frac{\hat{N}^{1/2}}{V}\sum_{kk'}\frac{\left(\hat{\Psi}_{kk'}^{\vpd}+\hat{\Psi}^{\dagger}_{-k,-k'}\right)e^{ik\cdot r+ik'\cdot(r+R)}}{2\sqrt{\omega_k\omega_{k'}}}.
    % \\    &+\frac{1}{V}\sum_{kk'}\frac{e^{ik\cdot r+ik'\cdot(r+R)}}{2\sqrt{\omega_k\omega_{k'}}}\bbraket{\delta\hat{\Psi}|\left(\hat{a}^\dagger_{-k'}\hat{a}_k+\hat{a}^\dagger_{-k}\hat{a}_{k'}\right)|\delta\hat{\Psi }}.
\end{equationS}
As in the quantum mechanical case, from this representation we can
\begin{itemize}
    \item Study the properties of the fluctuation field
    \begin{equation}
    \hat{D}_r=\bbrakettPsi{\hat{\phi}_r^2}-\hat{N}\braket{F|\hat{\phi}_r^2|F}.
    \end{equation}
    Its two-point function is given by 
    \begin{equation}
\left\langle\hat{D}_r\hat{D}_{r+R}\right\rangle=\frac{\hat{N}}{V^2}\sum_{kk'qq'}\frac{\left\langle\left(\hat{\Psi}_{kk'}^{\vpd}+\hat{\Psi}^{\dagger}_{-k,-k'}\right)\left(\hat{\Psi}_{qq'}^{\vpd}+\hat{\Psi}^{\dagger}_{-q,-q'}\right)\right\rangle e^{i(k+k')\cdot R}}{4\sqrt{\omega_k\omega_{k'}\omega_q\omega_{q'}}}.
\end{equation}
Because of translational invariance only momentum conserving processes will contribute ($k+k'=q+q'$). Using the resolvent formulas, this leads to
\begin{equation}
    \left\langle\hat{D}_r\hat{D}_{r+R}\right\rangle=-\frac{N}{2\lambda}\int _0^{\infty}\frac{ds}{\pi}\left[\frac{1}{V}\sum_P\frac{e^{iP\cdot{R}}}{1+B(-s^2,P)}\right],
\end{equation}
for $R\neq 0$. This is consistent with the fact that $1+B(-s^2,P)$ is proportional to the inverse propagator of the $\hat{\phi}^2$ composite field.
\item We could, in principle, compute corrections to $\bbrakettPsi{\hat{\phi}_r\hat{\phi}_{r+R}}$ to obtain, for example, the anomalous dimension $\eta$ in $d=2$. However, we would need to calculate the NNLO correction to the wavefunction. We can bypass this issue by looking instead at the operator
\begin{equation}
    \hat{\mathcal{O}}_k\equiv\bbrakettPsi{\omega_k\left(\hat{a}_k^\dagger\hat{a}_k+\frac{1}{2}\right)}=\frac{\omega_k\hat{N}}{2}+\omega_k\bbrakett{\delta\hat{\Psi}|\hat{a}_k^\dagger\hat{a}_k|\delta\hat{\Psi}},
\end{equation}
which does not suffer from this problem since $\hat{a}_k\ket{F}=0$. Furthermore
\begin{equation}\label{eqn:ONDescendant}
    \omega_k\left(\hat{a}_k^\dagger\hat{a}_k+\frac{1}{2}\right)=\frac{\hat{p}_k^\dagger\hat{p}_k}{2}+\frac{\omega_k^2\hat{\phi}_k^\dagger\hat{\phi}_k}{2},
\end{equation}
can be constructed via (lattice or time) differentiation of the field $\hat{\phi}_r$, which is a local operation, and can thus be interpreted as a descendant of $\hat{\phi}$. After some algebra (see Appendix~\ref{app:CritExp}), 
\begin{equationS}\label{eqn:ONDescendantII}
    \braket{\hat{\mathcal{O}}_k}&=\frac{\omega_k \hat{N}}{2}-\frac{2\lambda}{\pi V}\sum_q \int_0^{\infty}\frac{ds}{1+B(-s^2,k+q)}\left\{\frac{(\omega_k+\omega_q)^2-s^2}{\left[(\omega_k+\omega_q)^2+s^2\right]^2}\right\}\frac{\eta_{k-q}^2}{\omega_q},
\end{equationS}
In $d=2$, at $m=0$, small $|k|\ll 1$ and $\lambda \sim 1=\text{lattice scale}$, the non-analytic behaviour is given by
\begin{equation}
    \braket{\hat{\mathcal{O}}_k}_{\text{n.a.}}\approx \frac{\hat{N}|k|}{2}+\frac{4}{3\pi^2}|k|\log \left(\frac{|k|}{\upsilon}\right)+O(\hat{N}^{-1})\approx \frac{\hat{N}\upsilon}{2}\left(\frac{|k|}{\upsilon}\right)^{1+8/(3\pi^2 N)}, 
\end{equation}
where $\upsilon$ is a model-dependent constant. Thus, we recover the well-known anomalous dimension~\cite{Kleinert2001}
\begin{equation}
    \eta=\frac{8}{3\pi^2 N}.
\end{equation}
To be clear about the identification: the space-time propagator of $\hat{\phi}$ in $2+1$ dimensions is $k^{-2+\eta}$. To get the equal time correlator we have to integrate over frequencies, so it becomes $k^{-1+\eta}$ and the two derivatives in Eq.~(\ref{eqn:ONDescendant}) further transform this to $k^{1+\eta}$. Finally, since the $\hat{N}^{1/2}$ terms vanish for $\hat{\mathcal{O}}_k$, its fluctuations are also reduced, i.e. $(\delta \hat{\mathcal{O}}_k)\sim 1$ instead of $\sim \sqrt{N}$, as would be the case for any other generic field like $\hat{D}_r$.
\end{itemize}

\section{Finite temperature and models with fermions}\label{sec:FiniteTAndFermions}
The fluctuating $N$ formalism incorporates from the outset a restriction into permutationally symmetric states. Although this might not affect ground states, it does modify the weights of the states that contribute to thermal partition functions, and thus the finite temperature behaviour of the system in general. In this respect, the answers obtained using this formalism should be expected to differ from those obtained by calculating path integrals at finite imaginary time. On the other hand, $O(N)$ invariant states are a subset of permutationally symmetric states, so the formalism provides another way of studying the consequences of $O(N)$ invariance at finite temperature. We illustrate these points more concretely in the following sections, where we consider the non-interacting versions of both the QM and the $O(N)$ model ($\lambda=0$). We also include, at the end, a small discussion of models with fermions.
\subsection{Quantum Mechanical Model}\label{sec:QMFiniteT}
For completeness, we first discuss the finite temperature behaviour when only the permutationally symmetric constraint is enforced. We consider a modified version of the Hamiltonian in Eq.~(\ref{eqn:QMHami})
\begin{equation}
    \hat{H}_{\text{QM}}=\frac{1}{2}\sum_{\alpha}\left(\hat{p}_\alpha^2+\Omega^2\hat{x}_\alpha^2-
    \Omega\right)=\Omega\bbrakett{\hat{\Psi}|\hat{a}^\dagger\hat{a}|\hat{\Psi}},
\end{equation}
where we have introduced the frequency $\Omega$, $\hat{a}=(\Omega\hat{x}+i\hat{p})/\sqrt{2\Omega}$, and the $sp-$states $\ket{n}$ are indexed by the eigenvalues of $\hat{a}^\dagger\hat{a}$. This is essentially the Hamiltonian of a non-interacting BEC in a 1D harmonic trap, which is a problem that has led to many fruitful discussions regarding the differences and applicability of the various statistical mechanical ensembles~\cite{Grossman1996,HOLTHAUS1998198,KOCHAROVSKY2006291}. In the grand canonical ensemble, the number of species in $sp-$state $\ket{n}$ is given by
\begin{equation}
    N_n=\frac{1}{e^{\beta(\Omega n-\mu)}-1}.
\end{equation}
where $\mu<0$ is the chemical potential. Then
\begin{equation}
    N=\sum_{n}\frac{1}{e^{\beta(\Omega n-\mu)}-1}
\end{equation}
and the total energy is
\begin{equation}
    \braket{\hat{H}_{\text{QM}}}=\Omega\sum_{n}\frac{n}{e^{\beta(\Omega n-\mu)}-1}
\end{equation}
At high temperatures (high is to be defined shortly), the Bose-Einstein occupations can be replaced by the Maxwell-Boltzmann distribution, leading to
\begin{equation}
    \braket{\hat{H}_{\text{QM}}}=N\times\frac{\sum n e^{-\beta \Omega n}}{\sum e^{-\beta \Omega n}},
\end{equation}
which is just $N$ times the energy of the $N=1$ system. At lower temperatures the chemical potential approaches $0$ from below, $\beta\mu \to 0^-$, and the occupation of $n=0$ $sp-$state becomes macroscopic $\propto N$. The crossover from one behaviour to the other occurs when the total number of species cannot be accomodated in the $n\neq 0$ $sp-$states. In this regime, we need to work instead with the ``grand" canonical ensemble~\cite{KOCHAROVSKY2006291}, in which the $n=0$ $sp-$state acts as a reservoir of species for the excited $sp-$states, but the total species number is fixed (otherwise the resulting counting statistics of the $n=0$ $sp-$state are wrong). Within this treatment, the total occupation of the $n\neq 0$ states at $\beta\mu=0$ defines the crossover species number
\begin{equation}
    N_{\text{cross}}=\sum_{n=1}\frac{1}{e^{\beta \Omega n}-1},
\end{equation}
which must be large for the statistical description to make sense. This requires $\beta\Omega$ to be small, so
\begin{equation}
    N_{\text{cross}}\approx\frac{1}{\beta \Omega}\log\left(\frac{e^\gamma}{\beta\Omega}\right),
\end{equation}
where $\gamma$ is the Euler-Mascheroni constant. At a fixed large value of species number, setting $N=N_{\text{cross}}$ and solving for $\beta\Omega$ defines the regime of low vs. high temperatures, yielding $T_{\text{cross}}\approx \Omega N/\log(Ne^{\gamma})$. 
\subsubsection{O(N) constraint}
In the quantum mechanical case, the $O(N)$ constraint further reduces the admissible space of states that can contribute to the partition function. In fact, at a fixed $N$, the $O(N)$ invariant states are spanned by the set of orthogonal states
\begin{equation}
    \ket{l,N}=\bbrakett{\hat{\Psi}|(\hat{a}^\dagger)^2|\hat{\Psi}}^l\frac{(\hat{F}^\dagger)^N}{\sqrt{N!}}\ket{\text{vac}},
\end{equation}
where $\hat{F}=\brakett{0|\hat{\Psi}}$ destroys a species in the $\ket{n=0}$ $sp$-state. For completeness, we also include the fixed $N$ representation of the various terms that appear in $\ket{l,N}$ using the
\begin{equationS}
    \bbrakett{\hat{\Psi}|(\hat{a}^\dagger)^2|\hat{\Psi}}=\sum_{\nu=0}\sqrt{\nu+1}\sqrt{\nu+2}\hat{\Psi}_{\nu+2}^{\dagger}\hat{\Psi}_{\nu}^{\vpd}&\to \sum_{\alpha}^N\frac{(\Omega x_\alpha+\partial_{x_\alpha})^2}{2\Omega}\\
    \frac{(\hat{F}^{\dagger})^N}{\sqrt{N!}}\ket{\text{vac}}&\to\frac{e^{-\sum_{\alpha}x_\alpha^2/2}}{\pi^{N/4}}.
\end{equationS}
These states have energy $\hat{H}_{\text{QM}}\ket{l,N}=2\Omega l\ket{l,N}$, so the partition function of the system is
\begin{equation}
    \mathcal{Z}=\frac{1}{1-e^{-2\beta\Omega}},
\end{equation}
and is analytic everywhere. Nevertheless, we can still ask about condensation into the $sp-$ground state of the system $\ket{n=0}$. This means that we need to calculate the average value of $\hat{n}_0=\hat{F}^\dagger\hat{F}^{\vpd}/N$ in all the different $\ket{l,N}$ states. Repeatedly using the commutation relation
\begin{equation}
    \left[\bbrakett{\hat{\Psi}|\hat{a}^2|\hat{\Psi}},\bbrakett{\hat{\Psi}|(\hat{a}^\dagger)^2|\hat{\Psi}}\right]=\bbrakett{\hat{\Psi}|\left[\hat{a}^2,(\hat{a}^\dagger)^2\right]|\hat{\Psi}}=4\bbrakett{\hat{\Psi}|\hat{a}^\dagger\hat{a}|\hat{\Psi}}+2\hat{N}
\end{equation}
allows us to calculate the overlap of the $\ket{l,N}$ states, which is given by 
\begin{equation}
    \braket{l',N|l,N}= 2^{2l}l!\,\left(N/2\right)_{l}\,\delta_{ll'},
\end{equation}
where $(a)_l$ is the rising factorial. With this we can calculate the occupation of the $\ket{n=0}$ mode for each different $\ket{l,N}$ state
\begin{equation}
    \braket{\hat{n}_0}_{l}\equiv\frac{\braket{l,N|\hat{n}_0|l,N}}{\braket{l,N|l,N}}=\frac{\braket{l,N-1|l,N-1}}{\braket{l,N|l,N}}=\frac{(N/2-1/2)_{l}}{(N/2)_l}.
\end{equation}
At fixed $l$ and as $N\to \infty$ the ground state occupation fraction $\braket{\hat{N}_0}_{l,N}/N$ approaches $1$. For both $l$ and $N$ large we have that
\begin{equation}\label{eqn:FiniteTScalingCondensateFraction}
    \braket{\hat{n}_0}_{l}=\frac{1}{\sqrt{1+2l/N}}.
\end{equation}
We can obtain the occupation of the ground state in the canonical ensemble by doing a thermal sum
\begin{equation}
   \braket{\hat{n}_0}_{\beta}=\left(1-e^{-2\beta\Omega}\right)\sum_{l=1} e^{-2\beta\Omega l}\braket{\hat{n}_0}_{l},
\end{equation}
and we can use the asymptotic form for $\braket{\hat{n}_0}_{l}$ when $\beta\Omega\ll 1$, because then many large $l$ values contribute. In this regime we can do a continuum approximation, leading to
\begin{equation}
    \braket{\hat{n}_0}_{\beta}\approx 2\beta \Omega \int_0^{\infty}\frac{1}{\sqrt{1+2l/N}}e^{-2\beta\Omega l}\,dl=\sqrt{N\beta\Omega \pi}\,e^{N\beta\Omega}\text{erfc}\left(\sqrt{N\beta\Omega }\right),
\end{equation}
where $\text{erfc}$ is the complementary error function and the occupation becomes a function of the single variable $\zeta=\beta \Omega N$. When $\zeta\ll 1$ (low temperatures), the expression for $\braket{n_0}_\beta$ tends to $1$ as expected. When $\zeta\ll 1$, the occupation goes like $\sqrt{\zeta\pi}$, so that there is a crossover at temperatures of size $T\sim N\Omega$. Similarly, we can calculate the fluctuations in the occupation of the ground state
\begin{equation}\label{eqn:FiniteTGroundOccupationFluctuations}
    \braket{\hat{n}_0^2}_{l}=\frac{1}{1+2l/N}+O(N^{-1}),
\end{equation}
with thermal average
\begin{equation}
     \braket{\hat{n}_0^2}_{\beta}=N\beta\Omega\,e^{N\beta\Omega}\int_{N\beta\Omega}^{\infty}\frac{e^{-y}\,dy}{y}.
\end{equation}
In the microcanonical description, the fluctuations are suppressed at large $N$ because $\text{Var}(\hat{n}_0)_{l}$ goes like $N^{-1}$. In the canonical description, the fluctuations $\text{Var}(\hat{n}_0)_{\beta}$ go like $N^0$ so they are not suppressed at large $N$.

\subsection{O(N) model}
In the non-interacting case, the Hamiltonian of the $O(N)$ model can be written exactly as
\begin{equation}\label{eqn:nonInteractingOnModelFiniteT}
    \hat{H}^{O(N)}=\sum_k\omega_k\bbrakett{\hat{\Psi}|\hat{a}^\dagger_k\hat{a}_k|\hat{\Psi}},
\end{equation}
with $\omega_k^2=4\sum_{\delta r}\sin^2(k_{\delta r}/2)+m^2$ and $\beta=1/T$. The continuum limit is then defined by $m,T\ll 1$, so that the thermally occupied modes are those which vary on length scales much larger than the lattice spacing. To enforce conservation of $\hat{N}$ we work with the following state
\begin{equation}\label{eqn:FiniteT}
    \hat{\rho}=\exp\left[-\beta\left(\hat{H}^{O(N)}-\mu_N \bbrakett{\hat{\Psi}|\hat{\Psi}}\right)\right],
\end{equation}
where $\mu_N$ is a Lagrange multiplier. A given $sp-$state is characterized by $\{n_k\}$, i.e. the string of occupation numbers of each $sp$ $k$-mode, and is associated to one pair of creation and annihilation operators $\hat{\Psi}_{\{n_k\}}$, which create extended objects in that specific $sp-$state. The average number of those objects in the quantum state given by Eq.~(\ref{eqn:FiniteT}) is
\begin{equation}
    N_{\{n_k\}}=\left\langle\hat{\Psi}_{\{n_k\}}^\dagger\hat{\Psi}_{\{n_k\}}^{\vpd}\right\rangle=\left[e^{\beta\sum_k\left(\omega_k n_k\right)-\mu_N \beta}-1\right]^{-1}=\sum_{s=1}^\infty e^{s\beta\mu_N }e^{-s\beta\sum_k \omega_k n_k},
\end{equation}
which is just the standard Bose-Einstein distribution. The average number of species is then
\begin{equation}\label{eqn:FiniteTAverageN}
    \braket{\hat{N}}=\sum_{\{n_k\}}N_{\{n_k\}}=\sum_{s=1}^{\infty}\exp\left[s\beta \mu_N -\sum_{k}\log\left(1-e^{-s\beta\omega_k}\right)\right],
\end{equation}
while the average energy is
\begin{equation}
    \braket{\hat{H}}=\sum_{s=1}^{\infty}\exp\left[s\beta \mu_N -\sum_{k}\log\left(1-e^{-s\beta\omega_k}\right)\right]\sum_k\frac{\omega_k}{e^{s\beta\omega_k}-1}
\end{equation}
In the limit that $V T^2$ is very large, the contribution from $s=1$ dominates the sum (all other terms are smaller by factors of $e^{-VT^2}$ to some power), and the energy is given by
%and we have that
% \begin{equation}
%     \braket{\hat{N}}\approx e^{\beta \mu_N-\sum_k\log(1-e^{-\beta\omega_k}) }
% \end{equation}
% indicating that there will be a critical species number above which there will be Bose-Einstein condensation, obtained by setting $\mu_N=0$ and assuming that $m\neq 0$ (otherwise the number of $0$ energy $sp-$states is large enough to accomodate any number of species and condensation is hindered). In the non-condensed phase, the energy is given by
\begin{equation}
    \braket{\hat{H}}=\braket{\hat{N}}\sum_k\frac{\omega_k}{e^{\beta\omega_k}-1},
\end{equation}
which is just $N$ times the result for the $N=1$ system. The occupation of each $sp-$state becomes small too 
\begin{equation}
    \frac{N_{\{n_k\}}}{N}=\left[e^{\sum_k\log(1-e^{-\beta \omega_k })}\right]e^{-\beta\sum_k \omega_kn_k},
\end{equation}
where the prefactor is exponentially small in $VT^2$ and hence the occupations follow, to an excellent approximation, the Maxwell-Boltzman distribution. The free energy is given by
\begin{equation}
     F=\mu_N\braket{\hat{N}}-\frac{1}{\beta}\sum_{s=1}^{\infty}\frac{1}{s}\exp\left[s\beta \mu_N -\sum_{k}\log\left(1-e^{-s\beta\omega_k}\right)\right],
\end{equation}
and in the limit $VT^2\to \infty$ reduces to
\begin{equation}
    F=\frac{\braket{\hat{N}}}{\beta}\sum_k\log\left(1-e^{-\beta\omega_k}\right)+\frac{1}{\beta}\left(N\log N-N\right).
\end{equation}
Unlike the energy, there is a correction factor to the free energy. It has the same origin as the Gibbs correction factor that accounts for particle indistinguishability in statistical mechanics. In our case, this is due to permutational symmetry. As $VT^2\to\infty$ and for fixed $N$ then $\beta\mu\to-\infty$, making the Maxwell-Boltzmann distribution more and more accurate, and indicating that there are enough $sp-$states to accomodate a large number of species. Since, the number of excited $sp-$states is larger than in the associated quantum mechanical model (Section~\ref{sec:QMFiniteT}), a crossover from non-condensation to condensation should happen at a very large number of species (or equivalently at very small temperatures). When $m=0$, the $k=0$ $sp-$mode in Eq.~(\ref{eqn:nonInteractingOnModelFiniteT}) should be treated apart from the others since its contribution to $\hat{H}^{O(N)}$ is different, of the form $\hat{p}_0^2/2$.

 \subsubsection{O(N) constraint}
It is also possible to formulate the problem of finite temperature in the $O(N)$ symmetric sector. One possible route is to impose directly the $O(N)$ constraint. To leading order in $1/N$, this is equivalent to demanding that only the $\hat{\Psi}_{kk'}$ modes can be excited [as per Eq.~(\ref{eqn:ONconstraintLeadingOrder})] and thus we need to consider only the restricted Hamiltonian
\begin{equation}
    \hat{H}'=\frac{1}{2}\sum_{kk'}(\omega_k+\omega_{k'})\hat{\Psi}_{kk'}^\dagger\hat{\Psi}_{kk'}^{\vpd}.
\end{equation} 
This is still quadratic, so the partition function can be written down explicitly
\begin{align}\begin{split}
    \mathcal{Z}&=\mathrm{Tr}\left(e^{-\beta\hat{H}'}\right)=\exp\left[-\frac{1}{2}\sum_{k,k'}\log\left(1-e^{-\beta(\omega_k+\omega_{k'})}\right)\right],
\end{split}\end{align}
and the free-energy is
\begin{equation}
    -\beta F=-\frac{V^2}{2}\int\frac{d^2k\,d^2k'}{(2\pi)^4}\log\left(1-e^{-\beta(|k|+|k'|)}\right),
\end{equation}
where we have set $m=0$ and we work in the regime $VT^2\gg 1$ so that the thermal partition function is not sensitive to the shape of the underlying $2-$dimensional space. This integral can be calculated by first expanding the logarithm and performing the $k,k'$ integrals term by term, leading to
\begin{equation}
    -\beta F=\frac{V^2 T^4\zeta(5)}{8\pi^2}
\end{equation}
where $\zeta$ is the Riemann zeta function and in agreement with known results~\cite{shenker2011vectormodelssingletsector,Yoon2016-nc} [only after setting $V=4\pi$ and multiplying by $2$ because there is no double counting of momentum modes when the symmetry group is $U(N)$ instead of $O(N)$]. The associated energy is $\braket{\hat{H}}=V^2 T^5\zeta(5)/(2\pi^2)$, and the depletion of the condensed mode is
\begin{align}\begin{split}
    \frac{1}{2}\sum_{k,k'}\braket{\hat{\Psi}_{kk'}^\dagger\hat{\Psi}_{kk'}^{\vpd}}&=\frac{V^2}{2}\int \frac{d^2k d^2k'}{(2\pi)^4}\frac{1}{e^{\beta(|k|+|k'|)}-1}=\frac{V^2T^4\zeta(4)}{8\pi^2},
\end{split}\end{align}
indicating that condensation is lost at temperatures of size $VT^2\sim N^{1/2}$. As in the QM case, we should expect a crossover from condensation to non-condensation at such a temperature. Since this behaviour is not reflected in the free energy of the system, condensation is present only in so far as we take seriously the interpretation in terms of extended objects and ask questions about the occupation of given $sp-$states. 
%We can gain more intuition by computing further corrections to the condensate depletion using the $O(N)$ constraints [Eq.~(\ref{eqn:ONConstraint})] as in Section \ref{sec:HigherOrderQM}. This leads to (see Appendix~\ref{app:FiniteT2})
% \begin{equation}
%     \braket{\hat{N}-\hat{F}^\dagger\hat{F}}=\frac{V^2T^4\zeta(4)}{8\pi^2}+
% \end{equation}

We expect this phenomenon to be unrelated to the thermal phase transition found by Ref.~\cite{shenker2011vectormodelssingletsector} at the much higher temperature $VT^2\sim N$, which is a consequence of the existence of relations between the $\hat{\phi}_r$ variables when $N<VT^2$. We leave for future work a reproduction of the results from Ref.~\cite{shenker2011vectormodelssingletsector} within the formalism.
\subsection{Models with fermions}
Models with fermions can be analyzed in a similar fashion. Consider, for example, a lattice version of the Gross-Neveu model:
\begin{equationS}\label{eqn:fermionsHamiltonian}
    \hat{H}^{GN}&=\sum_{r=1}^L \sum_{\alpha=1}^N\Big(\hat{b}^\dagger_{r,\alpha}\hat{a}_{r+1,\alpha}+\hat{a}^\dagger_{r+1,\alpha}\hat{b}_{r,\alpha}-\hat{a}_{r,\alpha}^\dagger\hat{b}_{r,\alpha}-\hat{b}_{r,\alpha}^\dagger\hat{a}_{r,\alpha}\Big)\\
    &\hspace{6cm}-\frac{g^2}{2N}\sum_{r=1}^L\left[\sum_{\alpha=1}^N\Big(\hat{a}_{r,\alpha}^\dagger\hat{a}_{r,\alpha}^{\vpd}-\hat{b}_{r,\alpha}^\dagger\hat{b}_{r,\alpha}^{\vpd}\Big)\right]^2,
\end{equationS}
where $\alpha$ still indexes species, the model is defined on a one dimensional lattice of length $L$ and there are two types of fermions per lattice site per species: $\hat{a}_{r,\alpha}$ and $\hat{b}_{r,\alpha}$. The first sum in Eq.~(\ref{eqn:fermionsHamiltonian}) is the kinetic term, while the second sum is the interaction. We can rewrite this in the fluctuating $N$ Hilbert space as
    \begin{equation}
        \hat{H}^{\text{GN}}=\sum_{r=1}^L\bbra{\hat{\Psi}}\Big(\hat{b}^\dagger_{r}\hat{a}_{r+1}+\hat{a}^\dagger_{r+1}\hat{b}_{r}-\hat{a}_{r}^\dagger\hat{b}_{r}-\hat{b}_{r}^\dagger\hat{a}_{r}\Big)\kett{\hat{\Psi}}-\frac{g^2}{2\hat{N}}\sum_{r=1}^L\Big[\bbra{\hat{\Psi}}\Big(\hat{a}^\dagger_r\hat{a}_r-\hat{b}^\dagger_r\hat{b}_r\Big)\kett{\hat{\Psi}}\Big]^2,
    \end{equation}
where, importantly, $\kett{\hat{\Psi}}$ are still bosonic, even if the variables $\hat{a}_r$ and $\hat{b}_r$ are fermionic. This is because the bosonic commutation relations of $\kett{\hat{\Psi}}$ encode symmetry under permutation of species, irrespective of what the species are. We can then imagine that this describes extended one dimensional objects that have fermions living within them. We can rerun all the machinery of section~\ref{sec:O(N)model} and obtain the leading order ground state energy for small $m=g^2\braket{F|(\hat{a}_r^\dagger\hat{a}_r-\hat{b}^\dagger_r\hat{b}_r)|F}$ (as before, $\ket{F}$ is the state onto which the objects condense):
\begin{equation}
    \hat{H}^{GN}=\hat{N}\left[\frac{m^2}{2g^2}-\int_{-\pi}^{\pi}\frac{dk}{2\pi}\sqrt{4\sin^2(k/2)+m^2}\right]\approx \hat{N}\left[-\frac{2}{\pi}+\frac{m^2}{2g^2}+\frac{m^2}{4\pi}\log\left(\frac{m^2}{64e}\right)\right],
\end{equation}
with the self-consistent solution for $m$
\begin{equation}
    m=8e^{-\pi/g^2},
\end{equation}
the bound states with a mass on threshold ($m_{\text{bound}}=2m$), etc. This is in agreement with the known properties of the model. In this case no discussion of stability arises, but when $g^2<0$ the model can still be studied using the techniques of $\mathcal{PT}-$symmetry and adapted to other dimensionalities~\cite{Grable2023}.

\section{Summary and conclusions}
In this paper we have outlined the main features of a fluctuating $N$ approach to large $N$ vector models within the Hamiltonian framework. We began with a description of the method in the simpler context of quantum mechanics, where we introduced simplifying notation. We then applied the formalism to the $O(N)$ model, and analyzed the leading order and next-to-leading order contributions, reproducing known results for the ground state energy, correlators and critical exponents, while also showcasing the bi-local nature of the elementary excitations of the system. We finalized by discussing the application of the formalism to finite temperature systems and fermion models.

As mentioned in the main text, the nonlinear equation for the condensate wavefunction Eq.~(\ref{eqn:ONSelfConsistencyWavefunction}) has more than one solution, but we leave for future work the exploration of the nature of these other, highly energetic, states, which presumably correspond to solitons~\cite{universe10020099} [with spatially inhomogeneous $\sigma_r$ in Eq.~(\ref{eqn:ONspHamiltonian})]. Furthermore, this formalism can be adapted to consider quantum states that are antisymmetric with respect to permutations, so it would be interesting to study the properties of such states in the field theory context, where they would describe even more highly energetic excited states corresponding to a Fermi sea of extended objects. Finally, permutational symmetry can also be imposed in subsets of fields~\cite{Maslov199}, which opens the possibility of studying the effects of non-symmetric fluctuations in a controlled way. Such techniques could be employed to study, e.g., multi-flavor Schwinger models~\cite{COLEMAN1976,Dempsey2024}. 

\section{Acknowledgements}
The author thanks Paul Romatschke for igniting the interest in this approach to large $N$ vector models and for encouraging him to put out these results. The author thanks Paul Romatschke and Scott Lawrence for providing feedback on this manuscript. D.B. was supported by a Simons Investigator Award (Grant No. 511029) by the Simons collaboration on Ultra-Quantum Matter (UQM) which is funded by grants from the Simons Foundation (Grant No. 651440). We acknowledge additional funding support from the National Science Foundation under Grant Number PFC PHY-2317149  (Physics Frontier Center). D.B. acknowledges the hospitality of the KITP while parts of this work were completed. This research was supported in part by grant NSF PHY-2309135 to the Kavli Institute for Theoretical Physics (KITP).

\bibliographystyle{JHEP}
\bibliography{library}

\providecommand{\href}[2]{#2}\begingroup\raggedright\begin{thebibliography}{10}

\bibitem{klebanov2018tasi}
I.R.~Klebanov, F.~Popov and G.~Tarnopolsky, \emph{Tasi lectures on large $n$
  tensor models},  \href{https://arxiv.org/abs/1808.09434}{{\ttfamily
  1808.09434}}.

\bibitem{HOOFT1974}
G.~Hooft, \emph{A planar diagram theory for strong interactions},
  \href{https://doi.org/https://doi.org/10.1016/0550-3213(74)90154-0}{\emph{Nuclear
  Physics B} {\bfseries 72} (1974) 461}.

\bibitem{Maldacena1999}
J.~Maldacena, \emph{The large-n limit of superconformal field theories and
  supergravity},
  \href{https://doi.org/10.1023/A:1026654312961}{\emph{International Journal of
  Theoretical Physics} {\bfseries 38} (1999) 1113}.

\bibitem{Witten1998}
E.~Witten, \emph{Anti de sitter space and holography},
  \href{https://doi.org/https://link.intlpress.com/JDetail/1805563386639360001}{\emph{Adv.
  Theor. Math. Phys.} {\bfseries 2} (1998) 253}.

\bibitem{KLEBANOV2002213}
I.~Klebanov and A.~Polyakov, \emph{Ads dual of the critical o(n) vector model},
  \href{https://doi.org/https://doi.org/10.1016/S0370-2693(02)02980-5}{\emph{Physics
  Letters B} {\bfseries 550} (2002) 213}.

\bibitem{MOSHE2003}
M.~Moshe and J.~Zinn-Justin, \emph{Quantum field theory in the large n limit: a
  review},
  \href{https://doi.org/https://doi.org/10.1016/S0370-1573(03)00263-1}{\emph{Physics
  Reports} {\bfseries 385} (2003) 69}.

\bibitem{Maslov199}
V.P.~Maslov and O.Y.~Shvedov, \emph{Large-n expansion as a semiclassical
  approximation to the third-quantized theory},
  \href{https://doi.org/10.1103/PhysRevD.60.105012}{\emph{Phys. Rev. D}
  {\bfseries 60} (1999) 105012}.

\bibitem{Altland_Simons_2010}
A.~Altland and B.D.~Simons, \emph{Condensed Matter Field Theory}, Cambridge
  University Press, 2~ed. (2010).

\bibitem{Shvedov2001}
O.Y.~Shvedov, \emph{{Large-N theory from the axiomatic point of view}},
  \href{https://doi.org/10.1063/1.1387465}{\emph{Journal of Mathematical
  Physics} {\bfseries 42} (2001) 4197}.

\bibitem{Das2003}
S.R.~Das and A.~Jevicki, \emph{Large-n collective fields and holography},
  \href{https://doi.org/10.1103/PhysRevD.68.044011}{\emph{Phys. Rev. D}
  {\bfseries 68} (2003) 044011}.

\bibitem{Aharony2021}
O.~Aharony, S.M.~Chester and E.Y.~Urbach, \emph{A derivation of ads/cft for
  vector models}, \href{https://doi.org/10.1007/JHEP03(2021)208}{\emph{Journal
  of High Energy Physics} {\bfseries 2021} (2021) 208}.

\bibitem{Aharony2023b}
O.~Aharony, S.M.~Chester, T.~Sheaffer and E.Y.~Urbach, \emph{Explicit
  holography for vector models at finite n, volume and temperature},
  \href{https://doi.org/10.1007/JHEP03(2023)016}{\emph{Journal of High Energy
  Physics} {\bfseries 2023} (2023) 16}.

\bibitem{deMelloKoch2023}
R.~de~Mello~Koch, \emph{Gravitational dynamics from collective field theory},
  \href{https://doi.org/10.1007/JHEP10(2023)151}{\emph{Journal of High Energy
  Physics} {\bfseries 2023} (2023) 151}.

\bibitem{Dalmonte02072016}
M.~Dalmonte and S.~Montangero, \emph{Lattice gauge theory simulations in the
  quantum information era},
  \href{https://doi.org/10.1080/00107514.2016.1151199}{\emph{Contemporary
  Physics} {\bfseries 57} (2016) 388}.

\bibitem{berenstein2021bootstrappingsimpleqmsystems}
D.~Berenstein and G.~Hulsey, \emph{Bootstrapping simple qm systems},
  \href{https://arxiv.org/abs/2108.08757}{{\ttfamily 2108.08757}}.

\bibitem{Lawrence2023}
S.~Lawrence, \emph{Semidefinite programs at finite fermion density},
  \href{https://doi.org/10.1103/PhysRevD.107.094511}{\emph{Phys. Rev. D}
  {\bfseries 107} (2023) 094511}.

\bibitem{halimeh2023coldatomquantumsimulatorsgauge}
J.C.~Halimeh, M.~Aidelsburger, F.~Grusdt, P.~Hauke and B.~Yang, \emph{Cold-atom
  quantum simulators of gauge theories},
  \href{https://arxiv.org/abs/2310.12201}{{\ttfamily 2310.12201}}.

\bibitem{Bauer2023}
C.W.~Bauer, Z.~Davoudi, A.B.~Balantekin, T.~Bhattacharya, M.~Carena,
  W.A.~de~Jong et~al., \emph{Quantum simulation for high-energy physics},
  \href{https://doi.org/10.1103/PRXQuantum.4.027001}{\emph{PRX Quantum}
  {\bfseries 4} (2023) 027001}.

\bibitem{2006Leggett}
A.J.~Leggett, \emph{{Quantum Liquids: Bose condensation and Cooper pairing in
  condensed-matter systems}}, Oxford University Press (09, 2006),
  \href{https://doi.org/10.1093/acprof:oso/9780198526438.001.0001}{10.1093/acprof:oso/9780198526438.001.0001}.

\bibitem{RUBAKOV1988503}
V.~Rubakov, \emph{On third quantization and the cosmological constant},
  \href{https://doi.org/https://doi.org/10.1016/0370-2693(88)90108-6}{\emph{Physics
  Letters B} {\bfseries 214} (1988) 503}.

\bibitem{RoblesPerez2010}
S.~Robles-P\'erez and P.F.~Gonz\'alez-D\'{\i}az, \emph{Quantum state of the
  multiverse}, \href{https://doi.org/10.1103/PhysRevD.81.083529}{\emph{Phys.
  Rev. D} {\bfseries 81} (2010) 083529}.

\bibitem{Prosen_2008}
T.~Prosen, \emph{Third quantization: a general method to solve master equations
  for quadratic open fermi systems},
  \href{https://doi.org/10.1088/1367-2630/10/4/043026}{\emph{New Journal of
  Physics} {\bfseries 10} (2008) 043026}.

\bibitem{Ai2022}
W.-Y.~Ai, C.M.~Bender and S.~Sarkar, \emph{$\mathcal{P}\mathcal{T}$-symmetric
  $\ensuremath{-}g{\ensuremath{\varphi}}^{4}$ theory},
  \href{https://doi.org/10.1103/PhysRevD.106.125016}{\emph{Phys. Rev. D}
  {\bfseries 106} (2022) 125016}.

\bibitem{Romatschke2023}
P.~Romatschke, \emph{A solvable quantum field theory with asymptotic freedom in
  (3+1) dimensions},
  \href{https://doi.org/10.1142/S0217751X23501579}{\emph{International Journal
  of Modern Physics A} {\bfseries 38} (2023) 2350157}
  [\href{https://arxiv.org/abs/https://doi.org/10.1142/S0217751X23501579}{{\ttfamily
  https://doi.org/10.1142/S0217751X23501579}}].

\bibitem{romatschke2023quantum}
P.~Romatschke, \emph{Quantum field theory in large n wonderland: Three
  lectures},  \href{https://arxiv.org/abs/2310.00048}{{\ttfamily 2310.00048}}.

\bibitem{Gross1974}
D.J.~Gross and A.~Neveu, \emph{Dynamical symmetry breaking in asymptotically
  free field theories},
  \href{https://doi.org/10.1103/PhysRevD.10.3235}{\emph{Phys. Rev. D}
  {\bfseries 10} (1974) 3235}.

\bibitem{Ferrell1974}
R.A.~Ferrell and D.J.~Scalapino, \emph{Statistical mechanics of one-dimensional
  ginzburg-landau fields. ii. a test of the screening approximation
  ${n}^{\ensuremath{-}1}$ expansion)},
  \href{https://doi.org/10.1103/PhysRevA.9.846}{\emph{Phys. Rev. A} {\bfseries
  9} (1974) 846}.

\bibitem{Bray_1974}
A.J.~Bray, \emph{Statistical mechanics of one-dimensional ginzburg-landau
  fields: Feynman graph evaluation of the screening approximation (n-1
  expansion)}, \href{https://doi.org/10.1088/0305-4470/7/17/009}{\emph{Journal
  of Physics A: Mathematical, Nuclear and General} {\bfseries 7} (1974) 2144}.

\bibitem{Dolgov19880}
V.S.P.~A~. D.~Dolgov, V. L.~Eletskii, \emph{New approach to perturbation theory
  for a discrete spectrum (anharmonic oscillator)}, {\emph{Journal of
  Experimental and Theoretical Physics} {\bfseries 52} (1980) 861}.

\bibitem{Koudinov1982}
A.V.~Koudinov and M.A.~Smondyrev, \emph{1/n-expansion for the anharmonic
  oscillator}, \href{https://doi.org/10.1007/BF01596846}{\emph{Czechoslovak
  Journal of Physics B} {\bfseries 32} (1982) 556}.

\bibitem{Hioe1978}
F.~Hioe, D.~Macmillen and E.~Montroll, \emph{Quantum theory of anharmonic
  oscillators: Energy levels of a single and a pair of coupled oscillators with
  quartic coupling},
  \href{https://doi.org/https://doi.org/10.1016/0370-1573(78)90097-2}{\emph{Physics
  Reports} {\bfseries 43} (1978) 305}.

\bibitem{Susskind1964}
L.~Susskind and J.~Glogower, \emph{Quantum mechanical phase and time operator},
  \href{https://doi.org/10.1103/PhysicsPhysiqueFizika.1.49}{\emph{Physics
  Physique Fizika} {\bfseries 1} (1964) 49}.

\bibitem{Kromminga1962}
A.J.~Kromminga and M.~Bolsterli, \emph{Perturbation theory of many-boson
  systems}, \href{https://doi.org/10.1103/PhysRev.128.2887}{\emph{Phys. Rev.}
  {\bfseries 128} (1962) 2887}.

\bibitem{Moshe1983}
W.A.~Bardeen and M.~Moshe, \emph{Phase structure of the $\mathrm{O}(n)$ vector
  model}, \href{https://doi.org/10.1103/PhysRevD.28.1372}{\emph{Phys. Rev. D}
  {\bfseries 28} (1983) 1372}.

\bibitem{Giombi2020}
S.~Giombi, R.~Huang, I.R.~Klebanov, S.S.~Pufu and G.~Tarnopolsky, \emph{$o(n)$
  model in $4<d<6$: Instantons and complex cfts},
  \href{https://doi.org/10.1103/PhysRevD.101.045013}{\emph{Phys. Rev. D}
  {\bfseries 101} (2020) 045013}.

\bibitem{romatschke2024mass}
P.~Romatschke, C.-W.~Su and R.~Weller, \emph{Mass from nothing},
  \href{https://arxiv.org/abs/2405.00088}{{\ttfamily 2405.00088}}.

\bibitem{Abbott1976}
L.F.~Abbott, J.S.~Kang and H.J.~Schnitzer, \emph{Bound states, tachyons, and
  restoration of symmetry in the $\frac{1}{N}$ expansion},
  \href{https://doi.org/10.1103/PhysRevD.13.2212}{\emph{Phys. Rev. D}
  {\bfseries 13} (1976) 2212}.

\bibitem{ROMATSCHKE2023138270}
P.~Romatschke, \emph{What if $\phi^4$ theory in 4 dimensions is non-trivial in
  the continuum?},
  \href{https://doi.org/https://doi.org/10.1016/j.physletb.2023.138270}{\emph{Physics
  Letters B} {\bfseries 847} (2023) 138270}.

\bibitem{Abe1973}
R.~Abe and S.~Hikami, \emph{Breakdown of some scaling law relations in 1n
  expansion},
  \href{https://doi.org/https://doi.org/10.1016/0375-9601(73)90735-4}{\emph{Physics
  Letters A} {\bfseries 42} (1973) 419}.

\bibitem{ABE197311}
R.~Abe and S.~Hikami, \emph{Discontinuities of critical amplitude for specific
  heat},
  \href{https://doi.org/https://doi.org/10.1016/0375-9601(73)90482-9}{\emph{Physics
  Letters A} {\bfseries 45} (1973) 11}.

\bibitem{Aharony1974}
A.~Aharony, \emph{Scaling function for two-point correlations. ii. expansion to
  order $\frac{1}{n}$},
  \href{https://doi.org/10.1103/PhysRevB.10.2834}{\emph{Phys. Rev. B}
  {\bfseries 10} (1974) 2834}.

\bibitem{Wegner1972}
F.J.~Wegner, \emph{Corrections to scaling laws},
  \href{https://doi.org/10.1103/PhysRevB.5.4529}{\emph{Phys. Rev. B} {\bfseries
  5} (1972) 4529}.

\bibitem{Kleinert2001}
H.~Kleinert and V.~Schulte-Frohlinde, \emph{Critical Properties of
  $\phi^4$-Theories}, WORLD SCIENTIFIC (2001),
  \href{https://doi.org/10.1142/4733}{10.1142/4733}.

\bibitem{Grossman1996}
S.~Grossmann and M.~Holthaus, \emph{Microcanonical fluctuations of a bose
  system's ground state occupation number},
  \href{https://doi.org/10.1103/PhysRevE.54.3495}{\emph{Phys. Rev. E}
  {\bfseries 54} (1996) 3495}.

\bibitem{HOLTHAUS1998198}
M.~Holthaus, E.~Kalinowski and K.~Kirsten, \emph{Condensate fluctuations in
  trapped bose gases: Canonical vs. microcanonical ensemble},
  \href{https://doi.org/https://doi.org/10.1006/aphy.1998.5852}{\emph{Annals of
  Physics} {\bfseries 270} (1998) 198}.

\bibitem{KOCHAROVSKY2006291}
V.V.~Kocharovsky, V.V.~Kocharovsky, M.~Holthaus, C.~{Raymond Ooi},
  A.~Svidzinsky, W.~Ketterle et~al., \emph{Fluctuations in ideal and
  interacting bose–einstein condensates: From the laser phase transition
  analogy to squeezed states and bogoliubov quasiparticles},  vol.~53 of
  \emph{Advances In Atomic, Molecular, and Optical Physics}, pp.~291--411,
  Academic Press (2006),
  \href{https://doi.org/https://doi.org/10.1016/S1049-250X(06)53010-1}{DOI}.

\bibitem{shenker2011vectormodelssingletsector}
S.H.~Shenker and X.~Yin, \emph{Vector models in the singlet sector at finite
  temperature},  \href{https://arxiv.org/abs/1109.3519}{{\ttfamily 1109.3519}}.

\bibitem{Yoon2016-nc}
J.~Yoon, \emph{Finite temperature holography in higher spin theory/vector
  model}, Ph.D. thesis, Brown University, 2016.

\bibitem{Grable2023}
S.~Grable and M.~Weiner, \emph{A fully solvable model of fermionic interaction
  in 3 + 1d}, \href{https://doi.org/10.1007/JHEP09(2023)017}{\emph{Journal of
  High Energy Physics} {\bfseries 2023} (2023) 17}.

\bibitem{universe10020099}
A.~Jevicki, X.~Liu and J.~Zheng, \emph{Symmetries and the hilbert space of
  large n extended states},
  \href{https://doi.org/10.3390/universe10020099}{\emph{Universe} {\bfseries
  10} (2024) }.

\bibitem{COLEMAN1976}
S.~Coleman, \emph{More about the massive schwinger model},
  \href{https://doi.org/https://doi.org/10.1016/0003-4916(76)90280-3}{\emph{Annals
  of Physics} {\bfseries 101} (1976) 239}.

\bibitem{Dempsey2024}
R.~Dempsey, I.R.~Klebanov, S.S.~Pufu, B.T.~S\o{}gaard and B.~Zan, \emph{Phase
  diagram of the two-flavor schwinger model at zero temperature},
  \href{https://doi.org/10.1103/PhysRevLett.132.031603}{\emph{Phys. Rev. Lett.}
  {\bfseries 132} (2024) 031603}.

\end{thebibliography}\endgroup
\appendix
%%%%%%%%%%%%%%%%%%%%%%%%%%%%%%%%%%%%%%%%%%%%%%%%55
\section{Quantum Mechanical Model}
\subsection{Stationary points of the multispecies QM Hamiltonian}\label{app:QMStationaryPoints}
In this appendix we look for other solutions of Eq.~(\ref{eqn:QMSelfConsistency}):
\begin{equation}
    \bigg(\frac{\hat{p}^2}{2}+2\lambda\,\sigma\,\hat{x}^2\bigg)\ket{F}=\mu\ket{F}
\end{equation}
subject to the self-consistent relation $\sigma=\braket{F|\hat{x}^2|F}$. This is a harmonic oscillator $sp$-hamiltonian with frequency $\omega_{\sigma}=\sqrt{4\lambda \sigma}$, and its levels can be indexed by the excitation number $m$. In section~\ref{subsec:QMLeadingOrder} we chose $\ket{F}$ to be the $sp$-state with $m=0$ excitations. Here we consider other values of $m$. At level $m$, we have that $\mu_{\sigma}=\omega_{\sigma}(m+1/2)$, and thus 
\begin{equation}
    \sigma=\braket{m|\hat{x}^2|m}=\frac{(2m+1)}{2\omega_\sigma}.
\end{equation}
Plugging in the expression for $\omega_\sigma$ we solve for $\sigma$ as a function of $m$, leading to the following formulas
\begin{equationS}
    \sigma_m&=\frac{(2m+1)^{2/3}}{(16\lambda)^{1/3}}\\
    \omega_m&\equiv\omega_{\sigma_m}=(2\lambda)^{1/3}(2m+1)^{1/3}\\
    \mu_m&\equiv\mu_{\sigma_m}=\frac{(2\lambda)^{1/3}}{2}(2m+1)^{4/3}\\
    \hat{H}_m&=\hat{N}\bigg[\frac{3(2\lambda)^{1/3}}{8}(2m+1)^{4/3}\bigg].
\end{equationS}
There is one solution for each different $m$. The $m=0$ case leads to the lowest energy and is the solution written down in Eq.~(\ref{eqn:QMLOSolutions}). Importantly, the $sp$-state $\ket{m}$ is the $m$-th excited level of a harmonic oscillator with frequency $\omega_m$. Different $m$ have different frequencies, and so $\braket{m|m'}\neq 0$. Nevertheless, the associated quantum states in the multispecies Hilbert space are $\ket{m}^{\otimes N}$ and $\ket{m'}^{\otimes N}$, so that their overlap is $\braket{m|m'}^N$, which is exponentially small in $N$. 

We can also use the methods of Appendix~\ref{app:QMNLOHami} to find the NLO corrections about these other different solutions, akin to Eq~(\ref{eqn:QMNLODiagonal}). For a given $m$, the NLO corrections are (we quote only the final result)
\begin{equationS}
    \hat{H}_m^{\text{NLO}}/\omega_m&=\hat{N}\bigg[\frac{3}{8}(2m+1)\bigg]+\sum_{\nu\neq m}(\nu-m) \hat{\Psi}_{\nu,m}^\dagger \hat{\Psi}_{\nu,m}\\
    &+\bigg[\frac{\sqrt{(m+1)(m+2)}}{8(2m+1)}\Big(\hat{\Psi}_{m+2,m}e^{-i\hat{\varphi}}+\hat{\Psi}_{m+2,m}^\dagger e^{i\hat{\varphi}}\Big)\\&\hspace{3cm}+\frac{\sqrt{m(m-1)}}{8(2m+1)}\Big(\hat{\Psi}_{m-2,m}e^{-i\hat{\varphi}}+\hat{\Psi}_{m-2,m}^\dagger e^{i\hat{\varphi}}\Big)\bigg]^2,
\end{equationS}
where $\hat{\Psi}_{\nu,m}=\brakett{\nu,m|\hat{\Psi}}$ and $\ket{\nu,m}$ is the $\nu$-th excited level of $\hat{h}_{\sigma_m}$ (i.e. we fix $m$, but $\nu$ varies). Since $\hat{h}_{\sigma_m}$ changes with $m$, $\hat{\Psi}_{\nu,m}\neq \hat{\Psi}_{\nu,m'}$ unless $m=m'$ . Note that, at fixed $m$, the set $\{\hat{\Psi}_{1,m},\hat{\Psi}_{2,m},...\}$ is complete.

For each $m$ there is one such $\hat{H}_m$ that describe fluctuations about each of the different LO stationary solutions. For all $m\neq 0$, there are oscillators with negative frequency ($\nu<m$), indicating that only the $m=0$ solution is stable.

%%%%%%%%%%%%%%%%%%%%%%%%%%%%%%%%%%%%%%%%%%%%%%

\subsection{QM Hamiltonian to NLO}\label{app:QMNLOHami}
In this Appendix we begin with the exact Hamiltonian given in Eq.~(\ref{eqn:QMHami3}),
\begin{equation}\label{eqn:AppAQMHami}
    \hat{H}=\frac{1}{2}\bbrakett{\hat{\Psi}|\,\hat{p}^2|\hat{\Psi}}+\frac{\lambda}{\hat{N}} \bbrakett{\hat{\Psi}|\,\hat{x}^2|\hat{\Psi}}^2.
\end{equation}
and calculate the two leading terms in the $1/N$ expansion. We employ the decomposition
\begin{equation}\label{eqn:AppAStateODecomposition}
    \kett{\hat{\Psi}}=\ket{F}\hat{F}+\kett{\delta\hat{\Psi}},
\end{equation}
and the relation 
\begin{equation}\label{eqn:AppAN}
    \hat{N}=\hat{F}^\dagger\hat{F}+\bbrakett{\delta\hat{\Psi}|\delta\hat{\Psi}}.
\end{equation}
Since the mode associated to $\ket{F}$ is macroscopically populated, while the population of the rest of the modes is $\sim 1$, we will try to replace $\hat{F}^\dagger\hat{F}$ by $\hat{N}-\bbrakett{\delta\hat{\Psi}|\delta\hat{\Psi}}$ as much as possible. In particular, at certain points it will be useful to write $\hat{F}$ as
\begin{equation}\label{eqn:AppANumberPhase}
    \hat{F}=e^{i\hat{\varphi}}\,\sqrt{\big(\hat{F}^\dagger\hat{F}\big)}\approx  e^{i\hat{\varphi}}\hat{N}^{1/2}+O(N^{-1/2}),
\end{equation}
where $e^{i\hat{\varphi}}$ reduces $\hat{F}^\dagger\hat{F}$ by one with unit amplitude~\cite{Susskind1964}. Once we plug Eq.~(\ref{eqn:AppAStateODecomposition}) in the Hamiltonian Eq.~(\ref{eqn:AppAQMHami}), we will get many terms. To analyze them more easily, let us consider first the operators $\bbrakettPsi{\hat{x}^2}$ and $\bbrakettPsi{\hat{p}^2}$
\begin{equationS}
    \bbrakettPsi{\hat{x}^2}&=(\hat{F}^\dagger\hat{F})\braket{\hat{F}|\,\hat{x}^2|F}+\left[\hat{F}^\dagger\brakett{F|\,\hat{x}^2|\delta\hat{\Psi}}+\bbraket{\delta\hat{\Psi}|\,\hat{x}^2 |F}\hat{F}\right]+\bbrakett{\delta\hat{\Psi}|\,\hat{x}^2|\delta\hat{\Psi}}\\
    \bbrakettPsi{\hat{p}^2}&=(\hat{F}^\dagger\hat{F})\braket{\hat{F}|\,\hat{p}^2|F}+\left[\hat{F}^\dagger\brakett{F|\,\hat{p}^2|\delta\hat{\Psi}}+\bbraket{\delta\hat{\Psi}|\,\hat{p}^2 |F}\hat{F}\right]+\bbrakett{\delta\hat{\Psi}|\,\hat{p}^2|\delta\hat{\Psi}}.
\end{equationS}
The last two relations are exact, but now we replace all of the $\hat{F}$ to obtain approximately
\begin{equationS}
    \bbrakettPsi{\hat{x}^2}&\approx\hat{N}\braket{\hat{F}|\,\hat{x}^2|F}+\hat{N}^{1/2}\left[\brakett{F|\,\hat{x}^2|\delta\hat{\Psi}}e^{-i\hat{\varphi}}+\text{h.c.}\right]+\bbrakett{\delta\hat{\Psi}|\normOrd{\,\hat{x}^2}|\delta\hat{\Psi}}\\
    \bbrakettPsi{\hat{p}^2}&\approx\hat{N}\braket{\hat{F}|\,\hat{p}^2|F}+\hat{N}^{1/2}\left[\brakett{F|\,\hat{p}^2|\delta\hat{\Psi}}e^{-i\hat{\varphi}}+\text{h.c.}\right]+\bbrakett{\delta\hat{\Psi}|\normOrd{\,\hat{p}^2}|\delta\hat{\Psi}},
\end{equationS}
with corrections of size $O(\hat{N}^{-1/2})$ (not shown) and we have normal ordered the $sp$-operators with respect to the $sp$-vacuum $\ket{F}$, i.e. $\normOrd{\hat{O}}\equiv\hat{O}-\braket{F|\hat{O}|F}$ for any operator $\hat{O}$. After plugging these expression into the Hamiltonian we can group the resulting terms according to their power of $\hat{N}$, yielding:
\begin{equationS}
    \hat{H}_{\text{NLO}}&=\left[\frac{1}{2}\braket{F|\hat{p}^2|F}+\lambda\left(\braket{F|\,\hat{x}^2|F}^2\right)^2\right]\hat{N}^{1}\\
    &+\left[\brakett{F|\normOrd{\left(\frac{\hat{p}^2}{2}+2\lambda\hat{x}^2\braket{F|\,\hat{x}^2|F}\right)}|\delta\hat{\Psi}}e^{-i\hat{\varphi}}+\text{h.c.}\right]\hat{N}^{1/2}\\
    &+\left[\bbrakett{\delta\hat{\Psi}|\normOrd{\left(\frac{\hat{p}^2}{2}+2\lambda\hat{x}^2\braket{F|\,\hat{x}^2|F}\right)}|\delta\hat{\Psi}}+\lambda\left(\brakett{F|\,\hat{x}^2|\delta\hat{\Psi}}e^{-i\hat{\varphi}}+\text{h.c.}\right)^2\right]\hat{N}^{0}.
\end{equationS}
We note now that the first line of the previous equation is the LO approximation, whose value was found in Eq.~(\ref{eqn:QMLOSolutions}). The normal ordered operator in the second and third line corresponds to $\normOrd{\hat{h}_{\sigma_*}}$ defined in Eq.~(\ref{eqn:QMSelfConsistency}) and evaluated at the solution given in Eq.~(\ref{eqn:QMLOSolutions}). Hence, $\normOrd{\hat{h}_{\sigma_*}}=\hat{h}_{\sigma_*}-\mu$. Given that $\ket{F}$ is annihilated by this $sp$-operator, the second line vanishes. As a result, we have that
\begin{equation}
    \hat{H}_{\text{NLO}}\approx 0.375(2\lambda)^{1/3}\hat{N}+\bbrakett{\delta\hat{\Psi}|\left(\hat{h}_{\sigma_*}-\mu\right)|\delta\hat{\Psi}}+\lambda\left[\brakett{F|\,\hat{x}^2|\delta\hat{\Psi}}e^{-i\hat{\varphi}}+\text{h.c.}\right]^2
\end{equation}
which is Eq.~(\ref{eqn:QMNLOHami}) in the main text. Being more careful about 
\begin{equation}
    \hat{F}=e^{i\hat{\varphi}}\,\sqrt{\big(\hat{F}^\dagger\hat{F}\big)}=e^{i\hat{\varphi}}\,\left(\hat{N}-\bbrakett{\delta\hat{\Psi}|\delta\hat{\Psi}}\right)^{1/2}\approx e^{i\hat{\varphi}}\left(\hat{N}^{1/2}-\frac{\bbrakett{\delta\hat{\Psi}|\delta\hat{\Psi}}}{2\hat{N}^{1/2}}+...\right)
\end{equation}
leads to further corrections, with contributions from all orders in $1/N^{1/2}$ appearing in the Hamiltonian. For instance, the $\hat{N}^{-1/2}$ term is
\begin{align}\begin{split}
    \delta\hat{H}_{\text{NNLO}}=\frac{\lambda}{\hat{N}^{1/2}}\bigg\{\bbrakett{\delta\hat{\Psi}|\normOrd{\,\hat{x}^2}|\delta\hat{\Psi}}&\left[\brakett{F|\,\hat{x}^2|\delta\hat{\Psi}}e^{-i\hat{\varphi}}+\text{h.c.}\right]\\
    +&\left[\brakett{F|\,\hat{x}^2|\delta\hat{\Psi}}e^{-i\hat{\varphi}}+\text{h.c.}\right]\bbrakett{\delta\hat{\Psi}|\normOrd{\,\hat{x}^2}|\delta\hat{\Psi}}\bigg\}
\end{split}\end{align}
and the $N^{-1}$ correction is
\begin{equation}
    \delta\hat{H}_{\text{NNNLO}}=\frac{\lambda}{\hat{N}}\bigg[\bbrakett{\normOrd{\,\hat{x}^2}}^2-\frac{1}{2}\left\{\left(\brakett{F|\,\hat{x}^2|\delta\hat{\Psi}}e^{-i\hat{\varphi}}+\text{h.c.}\right),\left(\bbrakett{\delta\hat{\Psi}|\delta\hat{\Psi}}\brakett{F|\,\hat{x}^2|\delta\hat{\Psi}}e^{-i\hat{\varphi}}+\text{h.c.}\right)\right\}\bigg]
\end{equation}
where $\bbrakett{\normOrd{\,\hat{x}^2}}$ is shorthand for 
\begin{equation}
    \bbrakett{\delta\hat{\Psi}|\normOrd{\,\hat{x}^2}|\delta\hat{\Psi}}=\frac{1}{2(2\lambda)^{1/3}}\sum_{n=1}\left(2n\hat{\Psi}_n^\dagger\hat{\Psi}_n^{\vpd}+\sqrt{n}\sqrt{n-1}\hat{\Psi}_{n-2}^\dagger\hat{\Psi}_n^{\vpd}+\sqrt{n+1}\sqrt{n+2}\hat{\Psi}_{n+2}^\dagger\hat{\Psi}_n^{\vpd}\right)
\end{equation}
% while the $NNLO$ correction to the ground state wavefunction is given by
% \begin{equation}
%     \ket{\text{GND}}_{\text{NNLO}}=\ket{\text{GND}}_{\text{NLO}}+\left(\frac{1}{E_{\text{NLO}}-\hat{H}_{\text{NLO}}}\right)\ket{\text{GND}}_{\text{NLO}}
% \end{equation}

%%%%%%%%%%%%%%%%%%%%%%%%%%%%%%%%%%%%%%%%%%%%%%%%%%%%%

\subsection{O(N) constraint in QM model}\label{app:QMConstraint}
In this section we solve the $O(N)$ constraint to leading order in large $N$. As indicated in Section~\ref{sec:QMONInvariance}, $O(N) $ invariant states must satisfy
\begin{equation}
    \left[\big(\hat{x}_a\hat{p}_b-\hat{x}_b\hat{p}_a\big)\kett{\hat{\Psi}}_a\otimes\kett{\hat{\Psi}}_b\right]\ket{S}=0
\end{equation}
To proceed we express the $sp-$operator in terms of $\hat{a},\hat{b}$ defined as
\begin{equation}
    \hat{a}=\frac{\omega_{\sigma_*}\hat{x}_a+i\hat{p}_a}{\sqrt{2\omega_{\sigma_*}}},\hspace{1cm}
    \hat{b}=\frac{\omega_{\sigma_*}\hat{x}_b+i\hat{p}_b}{\sqrt{2\omega_{\sigma_*}}}
\end{equation}
and with $\omega_\sigma$ evaluated at $\sigma=\sigma_*$ given by Eq.~(\ref{eqn:QMLOSolutions}). Then the constraint operator is given by
\begin{equation}
    \left[\big(\hat{a}^\dagger\hat{b}^{\vpd}-\hat{b}^\dagger\hat{a}^{\vpd}\big)\kett{\hat{\Psi}}_a\otimes\kett{\hat{\Psi}}_b\right]
\end{equation}
Plugging in the decomposition $\kett{\hat{\Psi}}=\hat{F}\ket{F}+\kett{\delta\hat{\Psi}}$, with $\hat{F}\approx \sqrt{\hat{N}}e^{i\hat{\varphi}}$ and keeping only the terms $\propto \hat{F}^2\sim N$ and $\hat{F}\sim \sqrt{N}$ leads to the following contraint operators (we omit the tensor product symbols for notational simplicity)
\begin{equation}
    (\hat{a}^\dagger\hat{b}-\hat{b}^\dagger\hat{a})\ket{F}_a\ket{F}_b\,\hat{F}^2+\hat{F}\left(\hat{a}^\dagger\hat{b}-\hat{b}^\dagger\hat{a}\right)\left(\ket{F}_a\kett{\hat{\Psi}}_b+\kett{\delta\hat{\Psi}}_a\ket{F}_b\right)
\end{equation}
Since $\ket{F}_{a,b}$ is the vaccuum of $\hat{a},\hat{b}$, the first term evaluates to 0 and we are left with
\begin{equation}
    \hat{F}\left(\hat{b}\ket{1}_a\kett{\delta\hat{\Psi}}_b-\hat{a}\kett{\delta\hat{\Psi}}_a\ket{1}_b\right),
\end{equation}
where $\ket{1}_a=\hat{a}\ket{F}_a$ is the $sp-$state with $1$ excitation. Projecting the constraint onto $\bra{1}_a\bra{\nu-1}_b$ for $\nu=1,2...$ leads to 
\begin{equation}
    \hat{F}\left(\sqrt{\nu} \hat{\Psi}_{\nu}-\sqrt{2}\hat{\Psi}_2\delta_{\nu,2}\right).
\end{equation}
For $\nu=2$ this expresson is identically $0$, so it does not provide a condition. For $\nu\neq 2$ this expression is $\propto\hat{F}\hat{\Psi}_{\nu}$ and thus forces 
\begin{equation}
    \hat{\Psi}_\nu\ket{S}=0,\hspace{1cm}\nu\neq 2.
\end{equation} 
Thus, only the $sp$-states with $\nu=2$ can be occupied.

%%%%%%%%%%%%%%%%%%%%%%%%%%%%%%%%%%%%%%%%%%%%%%%%%%
\section{O(N) model}
\subsection{Numerical coefficients in the small $m$ expansion}\label{app:ONNumCoeff}
In this section we provide numerical values for the coefficients in the small $m$ expansion of
\begin{equation}
    \hat{H}^{O(N)}=\hat{N} V\sqrt{J_2}\bigg[\frac{1}{2}\int \frac{d^dk}{(2\pi)^d}\Big[4\sum_e\sin(k_e/2)^2+m^2\Big]^{1/2}-\frac{J_2^{3/2}}{16\lambda}\Big(m^2-\delta J\Big)^2\bigg],
\end{equation}
for $d=2$ and $d=3$. A naive Taylor expansion of the square root leads to divergent integrals, so we must be more careful. We illustrate the correct approach with the $d=2$ case. To begin with, we separate the integration region into two subregions: $|k|>\Lambda$ and $|k|<\Lambda$, where $\Lambda$ is an arbitrary cutoff that satisfies $m\ll\Lambda\ll 1$:
\begin{equationS}
    I_<\equiv\frac{1}{2}\int_{|k|<\Lambda} \frac{d^2k}{(2\pi)^2}\Big[4\sum_e\sin(k_e/2)^2+m^2\Big]^{1/2}\\
     I_>\equiv\frac{1}{2}\int_{|k|>\Lambda} \frac{d^2k}{(2\pi)^2}\Big[4\sum_e\sin(k_e/2)^2+m^2\Big]^{1/2}.
\end{equationS}
Since $\Lambda\ll 1$, the first integral can be simplified to
\begin{equation}
    I_<\approx\frac{1}{2}\int_{|k|<\Lambda} \frac{d^2k}{(2\pi)^2}\sqrt{|k|^2+m^2}=\frac{1}{12\pi}\left[\left(\Lambda^2+m^2\right)^{3/2}-|m|^3\right].
\end{equation}
Since $\Lambda\gg m$, we can expand the root to obtain
\begin{equation}
    I_<\approx \frac{\Lambda^3}{12\pi}+m^2\left(\frac{\Lambda}{8\pi}\right)-\frac{|m|^3}{12\pi}+m^4\left(\frac{1}{32\pi\Lambda}\right)+O(m^6).
\end{equation}
In the second integral $I_>$, we can safely Taylor expand the square root since $|k|>\Lambda\gg m$:
\begin{equationS}
    I_>&\approx m^0\left\{\frac{1}{2}\int_{|k|>\Lambda} \frac{d^2k}{(2\pi)^2}\Big[4\sum_e\sin(k_e/2)^2\Big]^{1/2}\right\}\\
    &+m^2\left\{\frac{1}{4}\int_{|k|>\Lambda} \frac{d^2k}{(2\pi)^2}\Big[4\sum_e\sin(k_e/2)^2\Big]^{-1/2}\right\}\\
    &+m^4\left\{-\frac{1}{16}\int_{|k|>\Lambda} \frac{d^2k}{(2\pi)^2}\Big[4\sum_e\sin(k_e/2)^2\Big]^{-3/2}\right\}+O(m^6)
\end{equationS}
We can now combine the integrals, collect terms with the same $m$ dependence and safely take the $\Lambda\to 0$ limit to obtain
\begin{equation}
    \frac{1}{2}\int \frac{d^2k}{(2\pi)^2}\left[4\sum_{\delta r}\sin(k_{\delta r}/2)^2+m^2\right]^{1/2}=A_2+B_2m^2-\frac{|m|^3}{12\pi}+C_2 m^4+O(m^6),
\end{equation}
where
\begin{equationS}%numerically checked that both analytical expressions give the same answer for A_2 and B_2
    A_2&=\frac{1}{2}\int \frac{d^2k}{(2\pi)^2}\left[4\sum_{\delta r}\sin^2(k_{\delta r}/2)\right]^{1/2}\approx 0.96\\%=2\int_{-\infty}^{\infty}\frac{ds}{\sqrt{2\pi}}e^{-2s^2}I_0(s^2)\left[I_0(s^2)-I_1(s^2)\right]\\&\\[8pt]
    B_2&=\frac{1}{4}\int \frac{d^2k}{(2\pi)^2}\left[4\sum_{\delta r}\sin^2(k_{\delta r}/2)\right]^{-1/2}\approx 0.16\\%=\frac{1}{4}\int_{-\infty}^{\infty}\frac{ds}{\sqrt{2\pi}}\left[e^{-s^2}I_0(s^2)\right]^2\\&\approx 0.16\\[8pt]
    C_2&=\lim_{\Lambda \to 0}\left\{\frac{1}{32\pi\Lambda}-\frac{1}{16}\int_{|k|>\Lambda}\frac{d^2 k}{(2\pi)^2}\left[4\sum_{\delta r}\sin^2(k_{\delta r}/2)\right]^{-3/2}\right\}%\\
    %&=\frac{1}{16}\int_{-\infty}^{\infty}\frac{ds}{\sqrt{2\pi}}\left\{\frac{1}{2\pi}-\left[s\, e^{-s^2}I_0(s^2)\right]^2\right\}\\
    \approx -0.0013.
\end{equationS}
are finite numbers. Similarly, for $d=3$ we have that
\begin{equationS}
    \frac{1}{2}\int \frac{d^3k}{(2\pi)^3}\left[4\sum_{\delta r}\sin(k_{\delta r}/2)^2+m^2\right]^{1/2}=A_3+B_3m^2&+\frac{m^4}{64\pi^2}\log\left(m^2\right)+C_3 m^4\\ &\hspace{1.8cm}+O(m^6\log m^2),
\end{equationS}
where
\begin{equationS}%numerically checked that both analytical expressions give the same answer for A_3 and B_3
    A_3&=\frac{1}{2}\int \frac{d^3k}{(2\pi)^3}\left[4\sum_{\delta r}\sin^2(k_{\delta r}/2)\right]^{1/2}\approx 1.2\\%=3\int_{-\infty}^{\infty}\frac{ds}{\sqrt{2\pi}}e^{-3s^2}\left[I_0(s^2)\right]^2\left[I_0(s^2)-I_1(s^2)\right]\\&\approx 1.2\\[8pt]
    B_3&=\frac{1}{4}\int \frac{d^3k}{(2\pi)^3}\left[4\sum_{\delta r}\sin^2(k_{\delta r}/2)\right]^{-1/2}\approx 0.11\\%=\frac{1}{4}\int_{-\infty}^{\infty}\frac{ds}{\sqrt{2\pi}}\left[e^{-s^2}I_0(s^2)\right]^3\\&\approx 0.11\\[8pt]
    C_3&=\lim_{\Lambda \to 0}\left\{\frac{1-4\log(2)}{128\pi^2}-\frac{\log(\Lambda)}{32\pi^2}-\frac{1}{16}\int_{|k|>\Lambda}\frac{d^3 k}{(2\pi)^3}\left[4\sum_{\delta r}\sin^2(k_{\delta r}/2)\right]^{-3/2}\right\}\\
    %&=-\frac{3}{128\pi^2}-\frac{3\log(2)}{64\pi^2}-\frac{\psi(1/2)}{64\pi^2}-\frac{1}{8}\int_{0}^{1}\frac{ds}{\sqrt{2\pi}}s^2\left[e^{-s^2}I_0(s^2)\right]^3\\
    %&\hspace{5.5cm}-\frac{1}{8}\int_{1}^{\infty}\frac{ds}{\sqrt{2\pi}}\left\{s^2\left[e^{-s^2}I_0(s^2)\right]^3-\frac{1}{2\sqrt{2}\pi^{3/2}s}\right\}\\
    &\approx-0.008
\end{equationS}

%%%%%%%%%%%%%%%%%%%%%%%%%%%%%%%%%%%%%%%%%%%%%%%%%%%%%%%%%5555

\subsection{Metastability for negative coupling}\label{app:ONMetastable}
Here we discuss the $O(N)$ model in the presence of a general potential $U(\phi^2)$, following Ref.~\cite{MOSHE2003}, to comment on the stability of the non-perturbatively generated minima within the fluctuating $N$ formalism. The Hamiltonian of the system is now
\begin{equation}\label{eqn:appOnHamiltonian}
    \hat{H}^{O(N)}=\sum_{r}\sum_{\alpha=1}^N\frac{(\hat{p}_r^\alpha)^2}{2}+\frac{1}{2}\sum_{r,\delta r} \sum_{\alpha=1}^NJ_{\delta r}^{\vphantom{\alpha}}\hat{\phi}_r^\alpha\hat{\phi}_{r+\delta r}^\alpha+N\sum_r U\left[\frac{1}{N}\sum_{\alpha=1}^N(\hat{\phi}_r^\alpha)^2\right],
\end{equation}
and in second quantized form
\begin{equationS}\label{eqn:appONHamiltonianSecondQuantized}
    \hat{H}^{O(N)}&=\frac{1}{2}\sum_{r}\bbra{\hat{\Psi}}\hat{p}_r^2\kett{\hat{\Psi}}+\frac{1}{2}\sum_{r,\delta r}J_{\delta r}\bbrakett{\hat{\Psi}|\hat{\phi}_r\hat{\phi}_{r+\delta r}|\hat{\Psi}}+\hat{N}\sum_r U\left[\frac{\bbra{\hat{\Psi}}\hat{\phi}_r^2\kett{\hat{\Psi}}}{\hat{N}}\right],
\end{equationS}
For $U(x)=\lambda x^2$ we recover the $\lambda\phi^4$ interaction. Running the same steps as in the main text, the $sp$-hamiltonian appearing in the optimization condition Eq.~(\ref{eqn:ONSelfConsistencyWavefunction}) is now
\begin{equationS}
    \hat{h}^{O(N)}=\frac{1}{2}\sum_{r} \hat{p}_r^2+\frac{1}{2}\sum_{r,\delta r} J_{\delta r}\hat{\phi}_r\hat{\phi}_{r+\delta r}+ \sum_{r} U'(\sigma_r) \hat{\phi}_r^2,
\end{equationS}
with $\sigma_r$ still defined as $\sigma_r=\braket{F|\hat{\phi}_r^2|F}$. The self-consistency condition, i.e. the gap equation, is now better expressed in terms of two equations for (homogeneous) $\sigma$ and $m$ (setting $J_2=1$):
\begin{equationS}
  \sigma&=\int \frac{d^d k}{2(2\pi)^d}\Big[4\sum_{\delta r}\sin(k_{\delta r}/2)^2+m^2\Big]^{-1/2}\\
  m^2&=2U'(\sigma)-\tilde{J}
\end{equationS}
From this representation of the gap condition, it becomes clear that $\sigma<\sigma_0$, where 
\begin{equation}
    \sigma_0=\int \frac{d^d k}{2(2\pi)^d}\Big[4\sum_{\delta r}\sin(k_{\delta r}/2)^2\Big]^{-1/2}
\end{equation}
which is a number of order $1$, and thus $m=0$ is a solution only when $\tilde{J}=\tilde{J}_c=U'(\sigma_0)$. At this massless point, we then have that
\begin{equationS}\label{eqn:app:gapEquations}
  \sigma_0-\sigma&=\int \frac{d^d k}{2(2\pi)^d}\left(\Big[4\sum_{\delta r}\sin(k_{\delta r}/2)^2\Big]^{-1/2}-\Big[4\sum_{\delta r}\sin(k_{\delta r}/2)^2+m^2\Big]^{-1/2}\right)\\[5pt]
  m^2&=2U'(\sigma)-2U'(\sigma_0).
\end{equationS}
We can look for another solution for small $m$ and hence small $\sigma_0-\sigma$. For definiteness let us choose $d=3$. To leading order we get
\begin{equationS}\label{eqn:app:TaylorExpandedGapEquation}
    \sigma_0-\sigma &\approx -\frac{m^2}{16\pi^2}\log m^2-m^2\left(\frac{1}{32\pi^2}+4C_3\right) \\[5pt]
    m^2&\approx 2U''(\sigma_0) (\sigma-\sigma_0),
\end{equationS}
where $C_3$ is still given in Appendix~\ref{app:ONNumCoeff}. The conditions $m^2>0$ and $\sigma<\sigma_0$ require that $U''(\sigma_0)<0$ for another solution to exist. In the case of $\lambda \phi^4$ this is the same as requiring that $\lambda<0$, but now this condition is just a local requirement on the general potential $U$. In fact, if we define $2\tilde{\lambda}=U''(\sigma_0)$, then the non-perturbative solution for $m$ is identical to Eq.~(\ref{eqn:ONdefm})
\begin{equation}
    \tilde{m}=\exp\left(\frac{2\pi^2}{\tilde{\lambda}}-\frac{1}{4}-32\pi^2 C_3\right)
\end{equation}
and the NLO Hamiltonian in Eq.~(\ref{eqn:NLOHamiltoniainONModel}) is modified by replacing $\lambda\to\tilde{\lambda}$. This is a good solution as long as the Taylor expansion of $U'(\sigma)$ about $\sigma_0$, invoked in Eq.~(\ref{eqn:app:TaylorExpandedGapEquation}) is valid. In other words, none of the Taylor expansion coefficients $\partial_{\sigma_0}^n U(\sigma_0)$ should be large enough to compensate for the smallness of $(\sigma-\sigma_0)^n$, which would require some fine-tuning of $U$. Consider now, as an example, the potential $U(x)=c_4x^2+c_6 x^3$, with $c_6>0$. This corresponds to a $c_4\phi^4+c_6\phi^6$ interaction that is bounded below. Then $\tilde{\lambda}=2c_4+6 c_6\sigma_0$ can be made small and negative by choosing $c_4$ adequately. At the same time, the potential leads to a symmetry broken absolute minimum with $m\sim 1$, a correlation length at the lattice scale, and an energy density that is parametrically larger by a factor $m^{-4}$.

%%%%%%%%%%%%%%%%%%%%%%%%%%%%%%%%%%%%%%%%%%%%%%%%%%%%%55555555555
\subsection{O(N) constraint in O(N) model}\label{app:ONConstraint}
In this section we formulate the constraint of $O(N)$ in the $O(N)$ model. In the first quantized formalism, the generator of rotations between species $\alpha$ and $\beta$ is
\begin{equation}
    \hat{R}_{\alpha\beta}^{O(N)}=\sum_r \left(\hat{\phi}^{\alpha}_r\hat{p}^{\beta}_r-\hat{\phi}^{\beta}_r\hat{p}_r^{\alpha}\right)=\sum_k\left(\hat{\phi}^{\alpha}_{k}\hat{p}^{\beta}_{-k}-\hat{\phi}^{\beta}_{k}\hat{p}_{-k}^{\alpha}\right),
\end{equation}
with $\hat{\phi}_k=V^{-1/2}\sum_r \hat{\phi}_r e^{-ikr}$ and $\hat{p}_k=V^{-1/2}\sum_r \hat{p}_r e^{-ikr}$. Switching to  annihilation operators $\hat{a}_k^{\alpha}=\Big(\omega_k\hat{\phi}^{\alpha}_k+ip_k^{\alpha}\Big)/\sqrt{2\omega_k}$ this becomes
\begin{equation}
    \hat{R}_{\alpha\beta}^{O(N)}=-i\sum_k\left[(\hat{a}^{\alpha}_k)^\dagger\hat{a}_k^{\beta}-(\hat{a}^{\beta}_k)^{\dagger}\hat{a}_k^{\alpha}\right].
\end{equation}
The $O(N)$ constraint is thus
\begin{equation}
    \left[\bbra{\hat{\Psi}}_a\otimes\bbra{\hat{\Psi}}_b\left(\sum_k\hat{a}_k^\dagger\hat{b}_k^{\vpd}-\hat{b}^\dagger_k\hat{a}^{\vpd}_k\right)^2\kett{\hat{\Psi}}_a\otimes\kett{\hat{\Psi}}_b\right]\ket{S}=0
\end{equation}
and factorizes to
\begin{equation}
    \left[\sum_k\left(\hat{a}_k^\dagger\hat{b}_k^{\vpd}-\hat{b}^\dagger_k\hat{a}^{\vpd}_k\right)\kett{\hat{\Psi}}_a\otimes\kett{\hat{\Psi}}_b\right]\ket{S}=0.
\end{equation}
%%%%%%%%%%%%%%%%%%%%%%%%%%%%%%%%%%%%%%%%%%%%%%%%%%%%%%%%%

\subsection{Bound state equations in the O(N) model}\label{app:ONBoundState}
In this appendix we derive the excitation energies by solving the eigenvalue problem for the matrix
\begin{equation}
     L_{Q,\tilde{Q}}^P=\left[(\omega_k+\omega_{k'})^2\delta_{Q\tilde{Q}}+\frac{4\eta_Q^{\vpd}\eta_{\tilde{Q}}\lambda}{V}\frac{\sqrt{(\omega_k+\omega_{k'})(\omega_q+\omega_{q'})}}{\sqrt{\omega_{k}\omega_{k'}\omega_{q}\omega_{q'}}}\right],
\end{equation}
where $k=(P+Q)/2$, $k'=(P-Q)/2$, $q=(P+\tilde{Q})/2$ and $q'=(P-\tilde{Q})/2$. The matrix $L^P$ has the structure of a diagonal matrix plus a rank one perturbation:
\begin{equation}
    L_{Q,\tilde{Q}}^P=\left[(\omega_k+\omega_{k'})^2\delta_{Q\tilde{Q}}+\frac{4\lambda}{V}v^P_Qv^P_{\tilde{Q}}\right],
\end{equation}
where $v^P_Q=\eta_Q \sqrt{\omega_k+\omega_{k'}}/\sqrt{\omega_{k}\omega_{k'}}$. Because of this, the eigenvectors $u_Q$ and eigenvalues $l$ satisfy
\begin{equation}
    (\omega_{k}+\omega_{k'})^2u_Q+\frac{4\lambda}{V}v_Q^P\sum_{\tilde{Q}}'v_{\tilde{Q}}^P u_{\tilde{Q}}=l\,u_Q\to u_Q=\frac{4\lambda v_Q^P}{V}\frac{\sum_{\tilde{Q}}v_{\tilde{Q}}^P u_{\tilde{Q}}}{l-(\omega_k+\omega_{k'})^2},
\end{equation}
where the prime in the sum indicates that only a restricted set of $\tilde{Q}$ should be considered to avoid double counting. Multiplying by $v_Q^P$, summing over $Q$ and cancelling common factors leads to
\begin{equation}
    -1=\frac{4\lambda}{V}\sum_{Q}\frac{(v_Q^P)^2}{(\omega_k+\omega_{k'})^2-l}=\frac{4\lambda}{V}\sum_{Q}'\left(\frac{1}{\omega_k}+\frac{1}{\omega_{k'}}\right)\frac{\eta_Q^2}{(\omega_k+\omega_{k'})^2-l}\equiv B(l,P),
\end{equation}
which defines the function $B(l,P)$. This is the same as Eq.~(\ref{eqn:BoundStateEquation}) in the main text. While $B(l,P)$ depends on the details of $\omega_k$ for generic $l$ and $P$, when $l,|P|\ll 1$, the function can be calculated as in Appendix~\ref{app:ONNumCoeff} by separating the $Q$ sum into two regions. We obtain
\begin{equationS}
        1+B(l,P)^{d=2}&=\frac{\lambda}{\pi}\left(-\frac{1}{\tilde{m}}+\frac{2m}{P^2-l^2}\int_0^{1}\frac{d\rho}{\rho^2+\frac{4m^2}{P^2-l^2}}\right)\\[8pt]
        1+B(l,P)^{d=3}&=\frac{\lambda}{2\pi^2}\left[\log\left(\frac{\tilde{m}}{m}\right)+\frac{1}{2}+\left(\frac{4m^2+P^2-l}{P^2-l}\right)\int_0^{1}\frac{dv}{v^2-\left(\frac{4m^2+P^2-l}{P^2-l}\right)}\right],
\end{equationS}
where $\tilde{m}$ are defined in Eq.~(\ref{eqn:ONdefm}) for $d=2$ and $d=3$. In the case of the bound state $P^2<l<P^2+4m^2$ so that direct integration of the expressions yields
\begin{equationS}
        1+B(l,P)^{d=2}&=\frac{\lambda}{\pi}\left[-\frac{1}{\tilde{m}}+\frac{1}{4m\theta}\log\left(\frac{1+\theta}{1-\theta}\right)\right]\\[8pt]
        1+B(l,P)^{d=3}&=\frac{\lambda}{2\pi^2}\left[\log\left(\frac{\tilde{m}}{m}\right)+\frac{1}{2}-\sqrt{\frac{1}{\theta^2}-1}\arctan\left(\frac{1}{\sqrt{\frac{1}{\theta^2}-1}}\right)\right],
\end{equationS}
where $\theta^2=(l-P^2)/(4m^2)$. Setting these expressions to $0$ and replacing $m\to \tilde {m}$ leads to the bound state equations in section~\ref{subsubsec:ONBoundStates}.
%%%%%%%%%%%%%%%%%%%%%%%%%%%%%%%%%%%%%%%%%%%%%%%%%%%%%%%%%%%%%%%%%%%%%%%%%%%55

\subsection{Correlator for critical exponent}\label{app:CritExp}
In this appendix we show how to calculate the expectation value of the operator
\begin{equationS}
    \hat{\mathcal{O}}_k&\equiv\bbrakettPsi{\omega_k\left(\hat{a}_k^\dagger\hat{a}_k+\frac{1}{2}\right)}=\frac{\omega_k\hat{N}}{2}+\omega_k\bbrakett{\delta\hat{\Psi}|\hat{a}_k^\dagger\hat{a}_k|\delta\hat{\Psi}}.
\end{equationS}
First we express it in terms of the $\hat{\Psi}_{kk'}$ modes:
\begin{equationS}
    \hat{\mathcal{O}}_k&=\frac{\omega_k\hat{N}}{2}+\omega_k\sum_{k'}\hat{\Psi}_{kk'}^\dagger\hat{\Psi}_{kk'}^{\vpd},
\end{equationS}
Then we express $kk'$ in terms of center-of-mass $P$ and momentum transfer $Q$ and switch to quadrature operators
\begin{equation}
    \hat{\mathcal{O}}_k=\frac{\omega_k\hat{N}}{2}+\frac{\omega_k}{2}\sum_{k'}\left(\hat{\Pi}_{P,Q}^\dagger\hat{\Pi}_{P,Q}^{\vpd}+\hat{X}_{P,Q}^\dagger\hat{X}_{P,Q}^{\vpd}-1\right).
\end{equation}
Averages can then be computed via resolvent methods such as 
\begin{align}\begin{split}
    \left\langle\hat{\Pi}_{P,Q}^\dagger\hat{\Pi}_{P,Q}^{\vpd}\right\rangle&=\frac{1}{\omega_k+\omega_{k'}}\left(\frac{\sqrt{L_P}}{2}\right)_{QQ}=\int_{\mathcal{C}_1} \frac{dz}{2\pi i}\frac{\sqrt{z}}{(\omega_k+\omega_{k'})}\frac{R_{QQ}^P(z)}{2}\\
    \left\langle\hat{X}_{P,Q}^\dagger\hat{X}_{P,Q}^{\vpd}\right\rangle&=(\omega+\omega_{k'})\left(\frac{1}{2\sqrt{L_P}}\right)_{QQ}=\int_{\mathcal{C}_1} \frac{dz}{2\pi i}(\omega_k+\omega_{k'})\frac{R_{QQ}^P(z)}{2\sqrt{z}},
\end{split}\end{align}
where
\begin{equationS}
    R_{QQ}^P(z)=\left(\frac{1}{L_P-z}\right)_{QQ}&=\frac{1}{(\omega_k+\omega_{k'})^2-z}\\
    &-\frac{1}{\big[(\omega_k+\omega_{k'})^2-z\big]^2}\times\frac{4\lambda  \eta_Q^2(\omega_k+\omega_{k'})}{V\omega_k\omega_{k'}}\times\frac{1}{1+B(z,P)}
\end{equationS}
and $k,k'$ are taken to be shorthand for $(P+Q)/2$ and $(P-Q)/2$ and $B(z,P)$ is defined in Eq.~(\ref{eqn:BoundStateEquation}). Summing over $Q$ and $P$ leads to Eq.~(\ref{eqn:Resolventt}). Plugging this into the equation for $\braket{\hat{\mathcal{O}}_k}$ leads to a cancellation of the $\lambda$ independent terms, and we are left with
\begin{equation}
    \braket{\hat{\mathcal{O}}_k}=\frac{\omega_k N}{2}-\frac{2\lambda}{V}\sum_{k'}\int_{\mathcal{C}_1}\frac{dz}{2\pi i}\frac{\eta_Q^2 (\omega_k+\omega_{k'})/\omega_{k'}}{1+B(z,k+k')}\times\frac{1}{2\big[(\omega_k+\omega_{k'})^2-z\big]^2}\times\left(\frac{\sqrt{z}}{\omega_k+\omega_{k'}}+\frac{\omega_k+\omega_{k'}}{\sqrt{z}}\right).
\end{equation}
Switching the integration contour to the negative $z$ axis and defining $z=-s^2$, we arrive at
\begin{equation}
    \braket{\hat{\mathcal{O}}_k}=\frac{\omega_k N}{2}-\frac{2\lambda}{V\pi}\sum_{k'}\int_0^\infty\frac{\eta_{k-k'}^2 \,ds/\omega_{k'}}{1+B(-s^2,k+k')}\times\frac{(\omega_k+\omega_{k'})^2-s^2}{\big[(\omega_k+\omega_{k'})^2-z\big]^2},
\end{equation}
which is Eq.~(\ref{eqn:ONDescendantII}) in the main text.

\end{document}